\documentclass[12pt, notitlepage]{article}

\usepackage[utf8]{inputenc}
\usepackage[a4paper,total={170mm,245mm}]{geometry}
\usepackage{fancyhdr}
\usepackage{mathrsfs}
\usepackage{amsmath}
\usepackage{graphicx}
\usepackage{subcaption}
\usepackage{hyperref}
\hypersetup{hidelinks,pdftitle=ReihsAttributables}
\usepackage{authblk}
\usepackage{tikz}
\usepackage[backend=bibtex, citestyle=numeric-comp, sorting=none, firstinits=true]{biblatex}

\newcommand{\Fig}[1]{Figure \ref{#1}}
\newcommand{\Tab}[1]{Table \ref{#1}}
\newcommand{\Sec}[1]{Section \ref{#1}}
\newcommand{\Equ}[1]{Equation \ref{#1}}

\title{Application of Attributables to the Correlation of Surveillance Radar Measurements}
\author[1,*]{B. Reihs}
\author[1]{A. Vananti}
\author[1]{T. Schildknecht}
\author[2]{J. A. Siminski}
\author[3]{T. Flohrer}
\affil[1]{Astronomical Institute University of Bern  (AIUB), Sidlerstrasse 5, 3012 Bern, Switzerland}
\affil[2]{IMS Space Consultancy, ESA/ESOC, Robert-Bosch-Str 5, 64293 Darmstadt, Germany}
\affil[3]{ESA/ESOC, Robert-Bosch-Str 5, 64293 Darmstadt, Germany}
\affil[*]{Corresponding Author: \href{mailto:benedikt.reihs@aiub.unibe.ch}{\underline{benedikt.reihs@aiub.unibe.ch}}}
\date{}

\begin{document}

	% Title
	\maketitle

\begin{abstract}

Space surveillance by radar is especially used for the low Earth orbit to maintain a database, also called catalogue, of objects on orbit. Among others, surveillance radars which are constantly scanning a region of interest in the sky are used for this purpose. The detections from such a radar which cannot be assigned to an already known catalogue object might not contain enough information to obtain a reliable initial orbit for a new catalogue entry from a single measured pass, also called tracklet. Instead, two tracklets can be combined to improve the quality of the initial orbit which leads to the correlation problem. This means that it has to be tested whether two tracklets belong to the same object and an initial orbit has to be derived by combining the tracklets. A common approach to condense the information in the tracklet is fitting them with so-called attributables. Because radar observations include different types of observables, the fitting of these attributables has to be considered as an important part of the entire correlation process. This paper analyses the effect of the attributable fitting considering the achieved accuracy and influence on the tracklet correlation. A new singularity-free coordinate system is introduced, which improves the results of the fitting and correlation. Finally, a test on a simulated survey scenario introduces two additional filters to remove false positive correlations. It is shown that the attributable-based approach can be applied successfully to tracklets of up to three minutes length with different detection frequencies.

\end{abstract}

\section{Introduction}

The ongoing growth of the space debris population is an increasing risk for operational satellites and the long-term sustainability of the near-Earth environment \cite{Liou340}. Thus, the importance of space surveillance is growing as it becomes necessary to maintain the orbits of space objects in a database, commonly referred to as catalogue. For the processing of observations, this creates two different tasks. The first one is the association of a sequence of measurements from a single pass, called a tracklet, with an object in the catalogue and its routine orbit improvement which is not considered here. The second association problem concerns the processing of the measurements which could not be associated with the catalogue, sometimes called uncorrelated tracks (UCT)\cite{hill2012comparison}. In these cases, an initial orbit determination from a single tracklet may not be sufficient to create a new catalogue object with a reliable orbit which leads to the topic of this work: the association of two tracklets in measurement space, called correlation in the following. Correlation methods using an orbit determination with a single tracklet can be found in e.g. \cite{hill2012comparison, vananti2017tracklet}, whereas the focus of the present work is the correlation based on an orbit determination using both tracklets combined as attributables.

Attributables \cite{milani2004orbit}, which are obtained by fitting a function over time to the raw measurements, have been introduced for optical observations and have been used successfully with different methods, e.g. \cite{siminski2014short, fujimoto2014association}. Also in case of radar measurements, attributables have already been used for initial orbit determination and correlation, e.g. \cite{demars2014probabilistic, gronchi2015computation, ma2018preliminary}. The authors of this paper also proposed a method based on the Lambert problem with a solution in the orbit space \cite{reihs2020method}. This method is used throughout the paper to derive the initial orbit and to perform the correlation. It uses two inertial positions which are derived from the radar measurements. From these two positions an initial orbit is derived under consideration of the secular J$_2$-perturbation. Depending on the time between the two positions, there might be several possible solutions with different numbers of revolutions between them. For each solution, the Mahalanobis distance $M_\text{d}$ \cite{mahadis} is calculated from the difference between the measured range-rates in the attributable and the calculated range-rate from the computed orbit which is scaled by the total uncertainty of the range-rates. This uncertainty is the sum of the measurement uncertainty and orbit uncertainty, obtained by a linear mapping of the observables which were used for the orbit determination onto the range-rate. The solution with the smallest $M_\text{d}$ is chosen as the most probable result and if it is smaller than a given threshold value $M_\text{d, thresh}$ the correlation is accepted. The value $M_\text{d, thresh}$ can be used for statistical gating, because it is $\chi$-distributed under the assumption of normally distributed errors.

Concerning the fitting of the attributables, \cite{maruskin2009correlation} described the least squares fitting of second order polynomials and their uncertainties for optical measurements, but so far there is no analysis on the realism of these fitted attributables, which is a prerequisite if they are used as an input for further processing, e.g. the described method. Thus this paper explores the accuracy of attributables for radar tracklets with regard to the estimated measurement value and its estimated uncertainty. Especially for the radar case there are up to four observables with three different physical interpretations, namely distance, velocity and direction angles.

Radars are predominantly used for space surveillance of the Low Earth Orbit (LEO). A surveillance radar uses the approach of regularly scanning a region of interest, called Field of Regard (FoR), using electric beamforming by phased-array radars \cite{saillant2015european, wilden2012low, gomez2019s3t}. Another possible operation mode is the creation of an additional beam to track a newly detected object \cite{wilden2017gestra, fonder2017fsy}, often referred to as stare-and-chase. Concerning surveillance radars, the size of the FoR and the time between two measurements of the same object during a pass can vary and thus also the input to the attributable fitting is not homogeneous. Previous work by \cite{mendijur2012management} also showed that surveillance radars are mainly useful for the regular observation of the space object population, whereas tracking radars are necessary to achieve high-fidelity orbits. The surveillance radars considered in this work are thus mainly aimed at providing regular observations.

In the following, the fitting of attributables is shortly reviewed in \Sec{sec:fits} considering radar observables, and different coordinate systems are discussed in \Sec{sec:acssys}. In \Sec{sec:simsce} the observation model for the simulations is introduced. This is used in \Sec{sec:attris} and \Sec{sec:corrs} to analyse the fitting of the attributables under various conditions and its effect on the correlation performance, before a complete processing of a simulated surveillance campaign is performed in \Sec{sec:survey}.

\section{Fitting Process}
\label{sec:fits}

The measurements are processed to form attributables at a reference epoch $t$ \cite{milani2004orbit}. The idea of the attributable is to average out noise and obtain a single, virtual measurement, which contains the condensed information of the entire tracklet. The attributable is obtained by fitting the raw measurements of each observable independently with a function over time and by picking the function values at a common epoch as the virtual measurements.

For this work, it is assumed that the radar provides four observables, namely the range $\rho$, the range-rate $\dot{\rho}$, the azimuth $az$ and the elevation $el$. Together with the reference epoch $t$ the attributable in this work is defined as: 
\begin{equation}
\label{eq:attri}
	\mathscr{A}_\text{AE} = \left\lbrace t, \rho, \dot{\rho}, az, el \right\rbrace \ ,
\end{equation}
which practically could also be extended to include additional information, e.g. the location of the observing station, but this is left out here for simplicity.

The fit to a specific observable's measurement vector $\vec{m}$ containing $M$ successive detections is calculated with an unweighted linear least squares approach for a polynomial of degree $n$ which contains $k = n+1$ parameters based on a Taylor series:
\begin{equation}
m(t) =  p_0 + p_1 \cdot \Delta t + \dots + \frac{p_n}{n!} \cdot \Delta t^n \ ,
\end{equation}
with the parameter vector  $\vec{p} = \left[ p_0, p_1, \dots , p_n \right]$. For the fitting process, all measurement epochs are used as the relative offset from the reference epoch which is set at the center of the measurement interval with $\Delta t=0$ s. The virtual measurements for the attributable are also set at the reference epoch, which has the advantage that the fitted parameter vector is equivalent to the values in the attributable vector without further processing, including the uncertainties and derivatives with the n-th derivative at $\Delta t=0$ being equal to $p_n$. For example, if the radar does not provide the range-rate as a separate measurement, it can be obtained from the first derivative of the range fit. This would also lead to a correlation of the estimated values and uncertainties for the range and range-rate, which may affect the tracklet correlation as well. The behaviour in such a case of correlated errors is analysed in \Sec{sec:cornoise}.

The solution of the fit is obtained by the classical least squares algorithm:
\begin{eqnarray}
\label{eq:uwlsq}
	\vec{p} & = & \left( A^T \cdot A \right)^{-1} \cdot \left( A^T \cdot \vec{m} \right) \ , \\
	A & = & \begin{bmatrix}
		1 & \Delta t_1 & \cdots &\frac{\Delta t_1^n}{n!}\\
		\vdots & \vdots & \ddots & \vdots\\
		1 & \Delta t_M & \cdots & \frac{\Delta t_M^n}{n!}\\
	\end{bmatrix} \ . 
\end{eqnarray}
The uncertainty $\sigma_{p_s}$ of the desired parameter $p_s, s \in \left[0, n\right]$ is derived from the covariance matrix $C = \left( A^T \cdot A \right)^{-1}$ \cite{bhattacharyya1977statistical}:
\begin{equation}
\label{eq:sigps}
	\sigma^2_{p_s} = \sigma^2_m \cdot C_{s,s} \ ,
\end{equation}
with $\sigma_m$ as the standard deviation of the measurements and $C_{s,s}$ the s-th value on the main diagonal of the covariance matrix. If $\sigma_m$ is unknown or shall be taken from the fit, which will be done throughout this paper, the residuals $\vec{r}$ of the fit can be used to obtain an unbiased estimate of the measurement noise with \cite{bhattacharyya1977statistical}: 
\begin{equation}
\label{eq:sigmeas}
\sigma^2_m = \frac{\vec{r} \cdot \vec{r}}{\left( M - k \right)}
\end{equation}

An extended fitting approach can be applied if the uncertainties of the observables are given by the sensor and they might even be correlated due to the measurement technique, making the use of \Equ{eq:sigmeas} unnecessary. In this case, the error covariance matrix of the measurements $\Sigma_M$ can be used for the generalised least squares:
\begin{equation}
\label{eq:gls}
	\vec{p}  =  \left( A^T \cdot \Sigma_M^{-1} \cdot A \right)^{-1} \cdot \left( A^T \cdot \Sigma_M^{-1} \cdot \vec{m} \right) \ .
\end{equation}
If the measurement errors are uncorrelated, i.e. $\Sigma_M$ has only elements on the main diagonal, this expression is equivalent to the weighted least squares. The covariance of the parameters is taken directly from the covariance matrix $\left( A^T \cdot \Sigma_M^{-1} \cdot A \right)^{-1}$ without further scaling using the residuals. This approach has two main possible sources of error. The first is that it requires a realistic input uncertainty of the measurements, otherwise the fit and the parameters' covariance will not be realistic. Additionally if different observables with correlated errors are fitted together, the correlations between the errors can become overly dominant and lead to unreasonable results especially for tracklets with a small number of data points. Throughout this paper, it is assumed that the measurement errors are unknown and thus \Equ{eq:uwlsq} is used.

\section{Applied Coordinate Systems}
\label{sec:acssys}

\begin{figure*}[htbp]
	\centering
	\subcaptionbox{Range}
	{\input{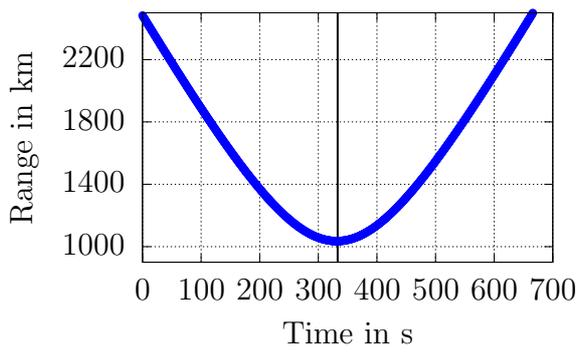}}%
	\hfill
	\subcaptionbox{Range-Rate}
	{\input{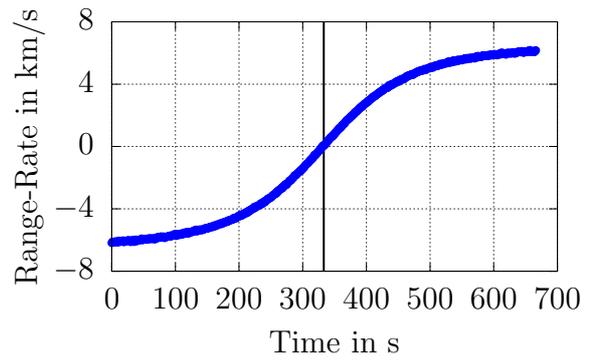}}%
	\vfill
	\subcaptionbox{Azimuth}
	{\input{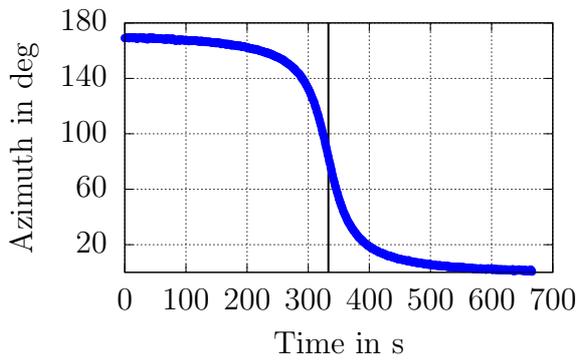}}%
	\hfill
	\subcaptionbox{Elevation}
	{\input{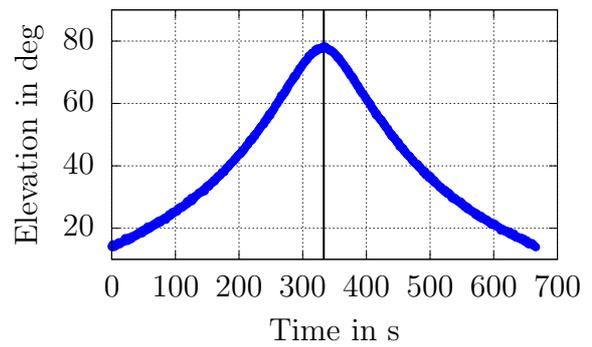}}%	
	\caption{Example graphs of radar observables over a full pass of an object in LEO. The vertical line indicates the time of closest approach.}
	\label{fig:obsall}
\end{figure*}

When attempting to fit a function to the measurements, the typical graph of the observables' function over the duration of a single pass has to be considered. Firstly, the four directly measurable radar observables as presented in the previous section are considered. \Fig{fig:obsall} depicts an example of the observables' development for a LEO object (semi-major axis $a \approx 7163$ km, eccentricity $e \approx 0.008$, inclination $i \approx 99^\circ$) over a nearly 10 minutes long radar pass. These graphs result from the orbital motion being projected into the observer's topocentric coordinate system. Although the specific numbers of the maximum and minimum values are depending on the geometry of each track, the general shape is similar for all passes. The vertical line in the plots marks the time of closest approach (TCA).
% Orbit: 7163.05,0.0076,99.09,55.40,112.22

The graph of the range has a parabola-like minimum at TCA, which is encompassed by two monotonic arcs towards larger values. Range-rate and azimuth are both close to the shape of the inverse tangent. For the range-rate the absolute values for minimum and maximum are nearly the same and it always switches from negative (approaching) to positive (departing). The extreme values of the azimuth are up to 180$^\circ$ apart depending on the maximum elevation of the pass, which also influences the slope at TCA. The higher this elevation, the more pronounced is the slope. For example if the object passes right above the observer, the azimuth would jump instantaneously by 180$^\circ$. The shape of the elevation is comparable to two exponential branches connected by a small parabola. From these plots, it can be concluded that it is presumably very difficult to fit even a higher order polynomial over longer time spans. Especially close to TCA, fitting the attributables may become problematic.

To overcome this issue, three options for deriving new observables from the given range-azimuth-elevation are introduced in the following with the goal to increase the robustness of the fitting process. All examples in the following show the same pass as in \Fig{fig:obsall}. The range-rate cannot be replaced and has to be used for all of the following systems.

\begin{figure*}[htbp]
	\centering
	\subcaptionbox{X-Coordinate}
	{\input{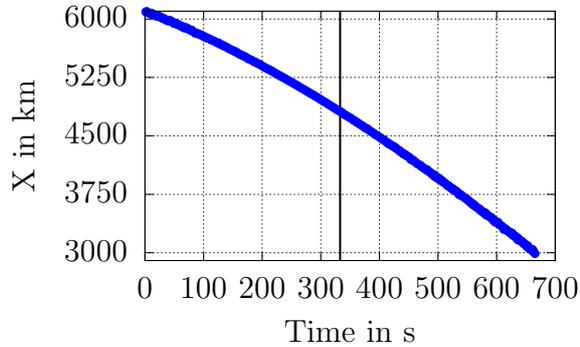}}%
	\hfill
	\vfill
	\subcaptionbox{Y-Coordinate}
	{\input{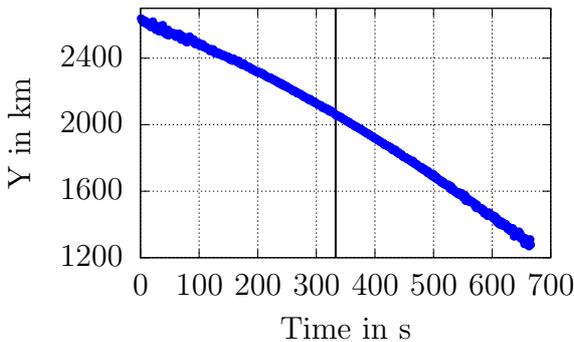}}%
	\hfill
	\subcaptionbox{Z-Coordinate}
	{\input{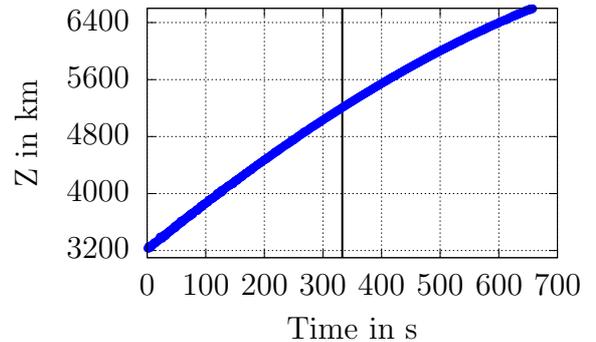}}%	
	\caption{Example graphs of geocentric inertial positions X, Y, Z over a full pass of an object in LEO. The vertical line indicates the time of closest approach.}
	\label{fig:obsxyz}
\end{figure*}

The first option is to transform each individual measurement to a position in a geocentric, inertial frame. This is possible by combining the ranges and angles of the radar measurement with the known location of the measurement station. The fit is then performed using the geocentric inertial positions, which can also be used as a direct input into the correlation method. This leads to the definition of an alternative attributable:
\begin{equation}
\label{eq:attrixyz}
\mathscr{A}_\text{XYZ} = \left\lbrace t, \dot{\rho}, X, Y, Z \right\rbrace \ .
\end{equation}
The two versions of the attributable are equivalent and can be inter-changed by an appropriate non-linear transformation, but performing the fit might be easier in one of the systems especially with regard to the uncertainties. The example plot using the geocentric inertial positions is shown in \Fig{fig:obsxyz}. In general, it can be seen that the three coordinates have a slightly curved but simple shape. The disadvantage of this system is that the behaviour of the individual components is heavily correlated with the orbit of the observed object and the location at which it is observed on the orbit. For example, observing a mid-inclination object at its most northern or southern point would lead to a parabola-like shape of the Z-coordinate, whereas the Z-coordinate would be a nearly straight line for a polar orbit. This reduces the consistency of observables over a population of many different objects. The uncertainties for each coordinate can be directly extracted from the fit neglecting the correlation of the errors after the transformation to the inertial coordinate system which will be discussed further at the end of this section.

\begin{figure*}[htbp]
	\centering
	\subcaptionbox{In-Plane Angle}
	{\input{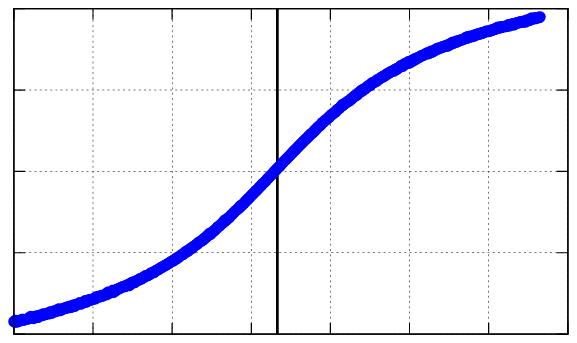}}%
	\hfill
	\subcaptionbox{Off-Plane Angle}
	{\input{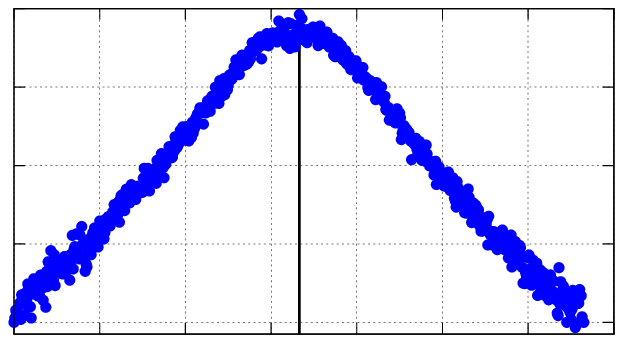}}%	
	\caption{Example graphs of topocentric AOS angles over a full pass of an object in LEO. The vertical line indicates the time of closest approach.}
	\label{fig:obsacstop}
\end{figure*}

The second alternative coordinate system is referred to in the following as topocentric Attributable Optimised Coordinate System (AOS). This system uses the same topocentric origin at the observer location and thus also the same range measurement as before. The only change is the definition of the angular observables. The new coordinate system is defined with the motivation to reduce the dependency of the angles on the pass geometry. It is defined by a reference plane through three points in the inertial space: the observer's location (neglecting its own motion), the first detection of the pass and the last detection of the pass. The angular position of each detection can thus be expressed as the in-plane angle $\theta_1$, which gives the direction angle from the station to the object within the reference plane comparable to the azimuth, and the off-plane angle $\theta_2$, which gives the elevation over the reference plane, also with the station as the reference. The direction $\theta_1 = 0^\circ$ can be defined arbitrarily within the plane, but here it is chosen to be consistent with the inertial direction of the vernal equinox projected onto the plane. The behaviour of these two angles can be seen in \Fig{fig:obsacstop} and they have a far lower variability among different passes compared to azimuth-elevation. These coordinates also do not have the singularity at $el=90^\circ$. The main influence is the length of the pass. For a short pass, the off-plane angle would be close to zero all the time and only long passes as in the example shown here lead to a distinct peak. To increase the comparability with the original topocentric frame and to use the measurements as an input for the correlation, the fitted AOS angles are transformed back to azimuth-elevation. The uncertainties which are estimated from the AOS fits have to be transformed as well and it has to be considered that this leads to a correlation of azimuth-elevation uncertainties which requires the use of a 2x2 covariance matrix with non-zero off-diagonal elements. The resulting attributable is the same as $\mathscr{A}_\text{AE}$. It should be noted that this coordinate system could also be used for optical observations by using unit vectors, because the ranges do not affect the definition of the plane.

\begin{figure*}[htbp]
	\centering
	\subcaptionbox{Geocentric Range}
	{\input{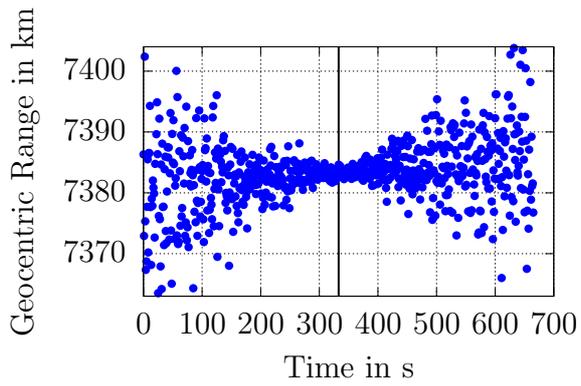}}%
	\hfill
	\subcaptionbox{Off-Plane Angle}
	{\input{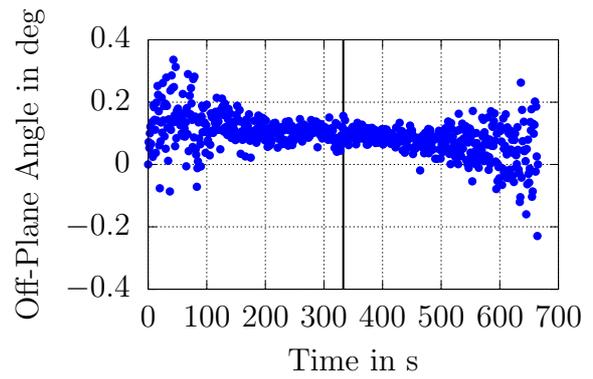}}%	
	\hfill
	\vfill
	\subcaptionbox{In-Plane Angle}
	{\input{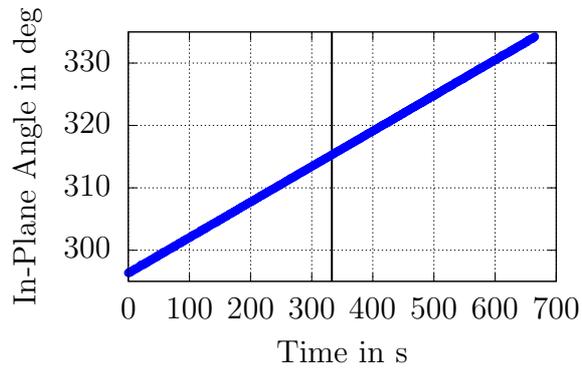}}%
	\caption{Example graphs of geocentric AOS observables over a full pass of an object in LEO. The vertical line indicates the time of closest approach.}
	\label{fig:obsacsgeo}
\end{figure*}

The last alternative is the geocentric AOS. The angles are the same as explained previously, but the observer's location is replaced as a reference point by the geocentric coordinate origin in the Earth centre. This shall reduce the variation in the off-plane angle further, which can be seen in \Fig{fig:obsacsgeo}. The newly defined plane is very close to the orbital plane. Thus the in-plane motion is approximately linear for nearly-circular LEO orbits and the off-plane angle should remain close to zero. In the example, due to the larger noise at larger ranges and the definition of the AOS plane at the edges of the tracklet, the off-plane angle $\theta_2$ is tilted against the expected $0^\circ$-line, but with a clear linear trend which would be equivalent to the orbital plane. Additionally, the topocentric range is replaced by the geocentric range for consistency and to facilitate the transformation into an inertial position, which is used as an input to the correlation routine. Again, this transformation leads to a statistical correlation of the errors in the inertial positions requiring a 3x3 covariance matrix in XYZ. The resulting attributable is the same as $\mathscr{A}_\text{XYZ}$. This geocentric AOS cannot be used with optical measurements because it requires the topocentric range information for the conversion to geocentric angles and range.

As mentioned in \Sec{sec:fits} all observables are fitted independently of each other because the amount of data is usually not sufficient for a fully correlated fit as presented in \Equ{eq:gls} and it is assumed that no initial information on the measurement accuracy is provided, which would be necessary for the error transformation into the new system. This refers to both versions of the AOS and the inertial system, where the errors of the transformed observables would be coupled due to the transformation from the azimuth-elevation system. The following results show that this approximation is sufficiently good for the purpose of correlation.

Finally, it should be remarked that if the errors in azimuth-elevation were very large (several degrees), the transformation to a geocentric system would lead to a breakdown of the errors' normal distribution due the curvature in the topocentric spherical coordinate system. However, it can be assumed that a radar sensor's angular uncertainty, i.e. the measurement errors in the next section, is small enough that this effect is not present.

\section{Simulation Scenarios}
\label{sec:simsce}

In order to make a realistic analysis of the fitting process, the lengths of the tracklets generated by a surveillance radar have to be estimated. For this, radar observations of the objects in the LEO population ($a < 8400$ km, $e < 0.1$, more than 10 000 objects in total), taken from Space-Track \cite{spacetrack}, have been simulated with a surveillance radar scanning an area of $60^{\circ} \times 20^{\circ}$, see \Tab{tab:noise}. The TLE are used to get a representative sample of space object orbits, while the actual propagation is performed using numerical propagation with \cite{orekit}. The dynamics are modelled with a 16x16 geopotential model, atmospheric drag (DTM2000 atmosphere, cannonball model), solar radiation pressure and lunisolar perturbations. The location of the radar is chosen at a low latitude to increase the coverage of orbital inclinations without simulating a specific real station, but this mainly influences the coverage of the observed satellites and should be of minor importance for the attributable fitting. The noise values given in the table increase with an increasing range due to the decrease of the signal-to-noise ratio \cite{skolnik2001introduction}. Additionally, the simulations can have three different intervals between the detections within a tracklet, denoted by 1f, 3f and 5f in the following corresponding to 1 s, 3 s, and 5 s between detections respectively. This is mostly dependent on the size of the FoR and the scanning strategy. A smaller FoR could be sampled at a higher rate but would produce shorter tracklets. This is a radar design trade-off.

\begin{figure}[htbp]
	\centering
	\input{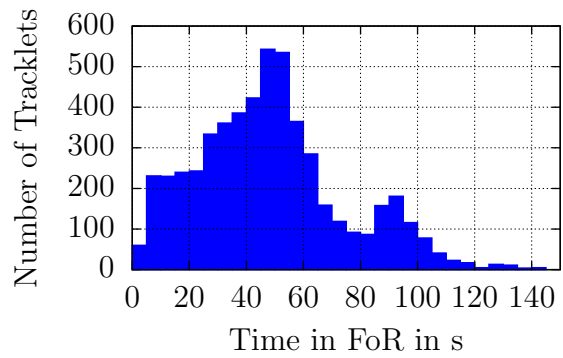}
	\caption{Dwell times in the field of regard for the simulated observation campaign.}
	\label{fig:dwell}
\end{figure}

The distribution of the overall dwell times in the FoR, and thus the effective tracklet lengths $t_T$, are shown in \Fig{fig:dwell}. It can be seen that the distribution peaks at approx. 50 s which is due to objects in sun-synchronous polar orbits ($a \approx 7200$ km) crossing the FoR in meridional direction. Another smaller peak occurs at approx. 100 s which is caused by objects in higher LEO regions ($a \approx 7800$ km). Depending on the location of the station, it may happen for a south-oriented FoR that an object with an inclination slightly smaller than the station's latitude passes exactly along the east-west direction through the FoR which can lead to a tracklet of several minutes. The following experiments will consider tracklet lengths up to three minutes (180 s) to cover at least a part of these outliers.

\begin{table}[htpb]
	\centering
	\caption{Radar sensor characteristics: field of regard and noise.}
	\begin{tabular}{|p{0.25\columnwidth} | p{0.2\columnwidth}|}
		\hline
		Radar, Latitude / Longitude & \centering 25$^\circ$N / 10$^\circ$E \tabularnewline
		\hline
		FoR, Azimuth & \centering 150$^\circ$ - 210$^\circ$ \tabularnewline
		\hline
		FoR, Elevation & \centering 50$^\circ$ - 70$^\circ$ \tabularnewline
		\hline
		Interval between detections & \centering 1 s, 3 s, 5 s \tabularnewline
		\hline
		Duration & \centering 24 hours \tabularnewline
		\hline
		\hline
		Reference range for noise & \centering 750 km \tabularnewline
		\hline
		Angles, $\sigma$ & \centering $0.17^\circ$ \tabularnewline
		\hline
		Range, $\sigma$ & \centering 20 m \tabularnewline
		\hline
		Rate, $\sigma$ & \centering 20 $\frac{\text{m}}{\text{s}}$ \tabularnewline
		\hline
	\end{tabular}
	\label{tab:noise}
\end{table}

\section{Results: Attributables}
\label{sec:attris}

\subsection{Evaluation Method}
\label{sec:atteval}

In the following, the evaluation process for the attributable fitting is explained. In this work we use only simulated radar measurements in order to have a reliable and precise ground truth for the observables. This is necessary to compare the fitting results and derive the statistical properties as described in the following.

\begin{figure*}[htbp]
	\centering
	{\input{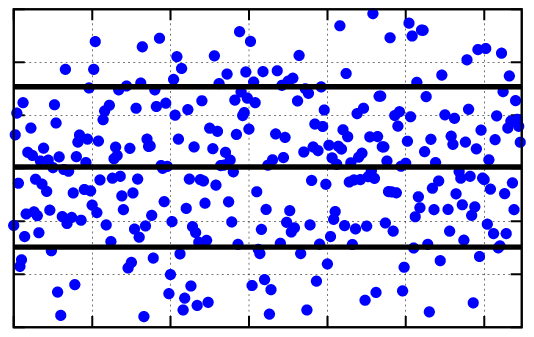}%
		\hfill
		\input{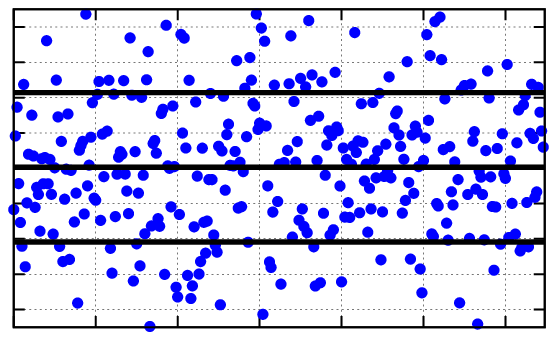}}%
	\caption{Example of absolute and relative errors: azimuth. The bold lines indicate the mean and mean $\pm$ standard deviation}
	\label{fig:exaerror}
\end{figure*}

Different experiments concerning the fitting of the attributables are covered in this paper. For these tests, a subpopulation of the previously introduced LEO population is used consisting of 1500 objects. As the location of the FoR is not important for the following tests, the restrictions on the visibility considering the FoR are reduced to $el > 5^\circ$ and $\rho < 2500$ km to maximise the number of obtained tracklets which usually results in 5000 - 6000 tracklets per experiment. For this analysis, the concept of absolute and relative errors is introduced regarding the observables. The absolute error $\Delta A_{x}$ of observable $x$ is derived by taking the difference between the value obtained via the attributable fitting and the true, noiseless value of the observable at the same epoch. It is important that the mean of the absolute error $\mu_{A,{x}}$ over the entire population is close to zero, otherwise the estimation of the parameter is biased. If the absolute error is related to the estimated uncertainty $\sigma_{x}$, the relative error for a single value is calculated as $\Delta R_{x} = \frac{\Delta A_{x}}{\sigma_{x}}$ which transforms the error to the z-scale, a normalised Gaussian distribution with the idealised properties of the mean $\mu = 0$ and the standard deviation $\sigma = 1$, to make the different individual errors comparable. Actually, the resulting distribution would be a Student's t-distribution, but for a sufficient number of data points, this distribution converges to a standard normal distribution which will be assumed for all following experiments. Thus the overall standard deviation $\sigma_R$ of the relative error over the entire population of attributables should be close to one to confirm a realistic estimation of the error. This value is calculated in two steps. After getting the initial standard deviation $\sigma_{R,0}$, all values with $\Delta R_{x} > 3\cdot \sigma_{R,0}$ are removed to avoid a biased statistic due to extreme outliers and the final result $\sigma_R$ is obtained from the new data set. The same approach is also applied to the absolute errors. The absolute and relative errors of a population are shown in \Fig{fig:exaerror} with the azimuth angle as an example. The plots in the following sections will condense this information to an errorbar showing $\mu \pm \sigma$ for both the absolute and relative error.

For the two AOS, it has been mentioned in \Sec{sec:acssys} that the errors between the observables are correlated in the attributable after the transformation back to azimuth-elevation or inertial coordinates, respectively. To check whether these correlations are also well-captured, the squared Mahalanobis distance between the fitted attributable and the noiseless reference measurement is calculated \cite{mahadis}:
\begin{equation}
M_d^2 = \vec{d_M}^{\ T} \cdot C_{\text{meas}}^{-1} \cdot \vec{d_M} \ ,
\end{equation}
where $\vec{d_M}$ is the vector containing the differences between the fitted value and the ground truth. $C_{\text{meas}}$ is the estimated covariance matrix. Theoretically $M_d^2$ should be distributed according to a $\chi^2_\nu$-distribution with $\nu$ degrees of freedom. This distribution has a mean of $\mu = \nu$ and a variance of $\sigma^2 = 2\nu$ \cite{abramowitz1965handbook}, which can be checked against the values from the fitted population. For the cases here, the transformation from topocentric AOS to azimuth-elevation leads to a 2x2 covariance matrix for the angles and thus $\nu=2$ and from geocentric AOS to inertial position gives $\nu=3$. The other fits are independent and thus not checked together with the correlated errors. 
%  (Reference! \url{http://people.math.sfu.ca/~cbm/aands/page_943.htm})
	
\subsection{Length of Tracklet}

The first experiment is performed to analyse the influence of the tracklet length on the accuracy of the fitting process for the observables. This has been performed with all introduced measurement frequencies but it was found that, although it causes slight differences, this does not change the overall trend of the accuracies and thus only one selected frequency value per observable is shown as an example. The following plots compare different polynomials, up to the fourth order. Higher orders will be investigated in \Sec{sec:order}.

\subsubsection{Topocentric Range and Range-Rate}

\begin{figure*}[htbp]
	\centering
	{\input{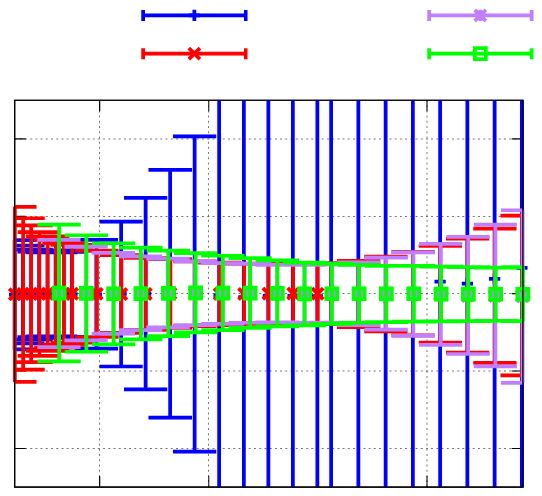}%
		\hfill
		\input{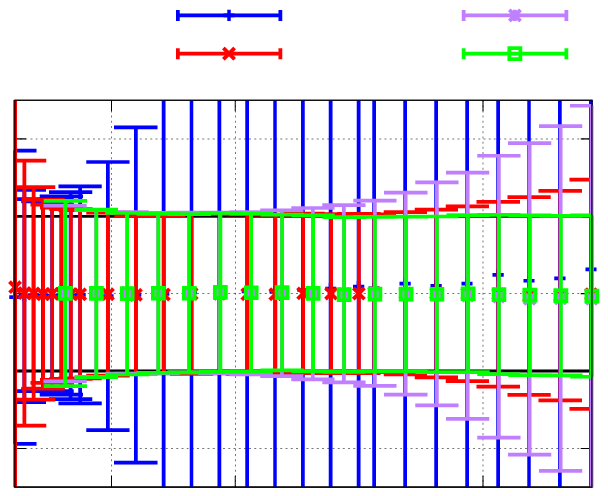}}%
	\caption{Range-rate errors for different tracklet lengths using 3f-data.}
	\label{fig:lenrate}
\end{figure*}

The range-rate has to be fitted for all coordinate systems and is used for the correlation decision \cite{reihs2020method}. An example of the fit results for the first four orders is shown in \Fig{fig:lenrate}. It can be seen that the point at which the quadratic fit becomes better than the linear one considering the absolute error is at approx. 30 s and after approx. 130 s the fourth order fit becomes the best. Also for the relative errors, the given orders of the polynomial are close to one within their respective intervals, which indicates a well-approximated uncertainty.

\begin{figure*}[htbp]
	\centering
	{\input{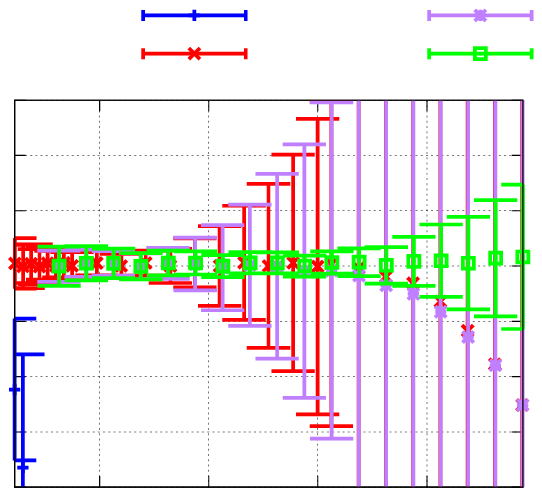}%
		\hfill
		\input{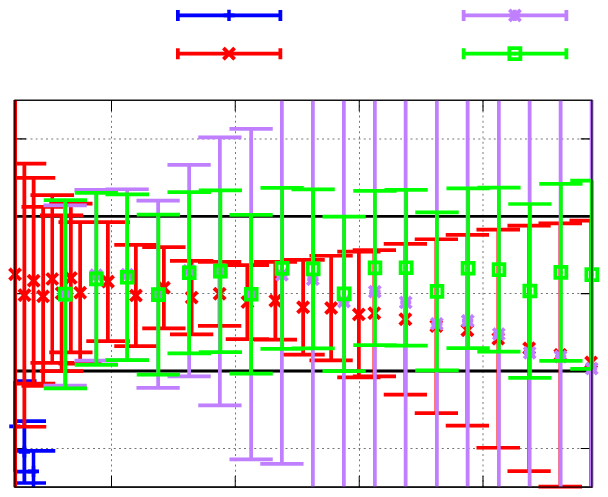}}%
	\caption{Topocentric range errors for different tracklet lengths using 3f-data.}
	\label{fig:lenrho}
\end{figure*}

\Fig{fig:lenrho} depicts the errors of the topocentric range, which is required for the two topocentric coordinate systems. Even for relatively short tracklets, the linear fit introduces a negative bias due to the curvature of the range measurements and is not even in the plotted range any longer. Independent from the measurement frequency, this bias reaches approximately -1 km at 30 s and grows exponentially to -10 km at 100 s. Thus only the quadratic fit is a reasonable choice for the range attributable from the beginning, but its absolute error is growing for tracklets longer than 60 s. Afterwards a fourth order fit is the better choice, which has increased errors from approx. 140 s onwards. The relative error for the quadratic fit is close to one for lengths smaller than 40 s. The fourth order fit shows some inconsistencies regarding the relative error, but it is a much better approximation than the quadratic one after approx. 50 s.

\subsubsection{Azimuth-Elevation}

\begin{figure*}[htbp]
	\centering
	{\input{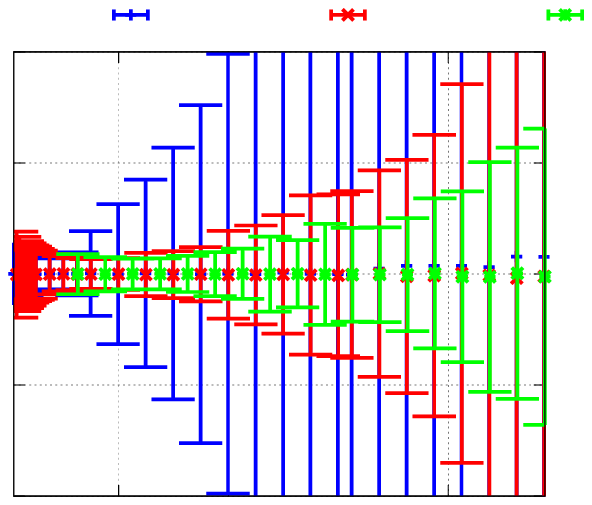}%
		\hfill
		\input{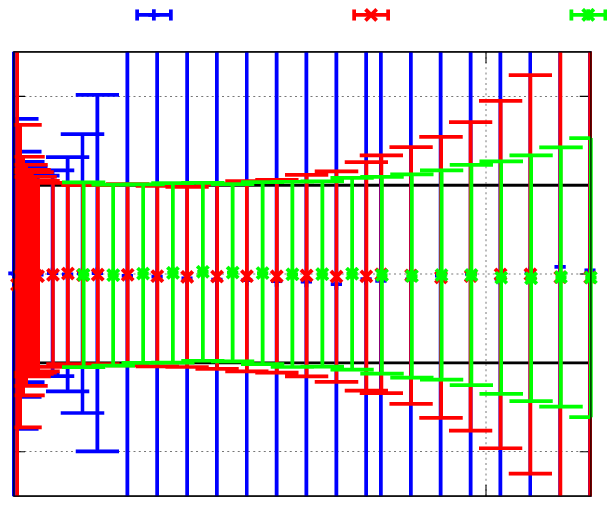}}%
	\caption{Azimuth errors for different tracklet lengths using 1f-data.}
	\label{fig:lenazi}
\end{figure*}

The result of the azimuth fitting is shown in \Fig{fig:lenazi}. At approximately 25 s, the absolute error of the quadratic fit becomes lower than the linear one's which increases rapidly after this time. Regarding the relative error, both fits are comparable until the same 25 s mark and afterwards the linear fit has larger deviations than one which implies an underestimation of the uncertainty in the fitting process. The quadratic fit shows a good representation of the error up to approx. 80 s. Afterwards the fourth order fit has the smaller absolute error and the better error distribution.

\begin{figure*}[htbp]
	\centering
	{\input{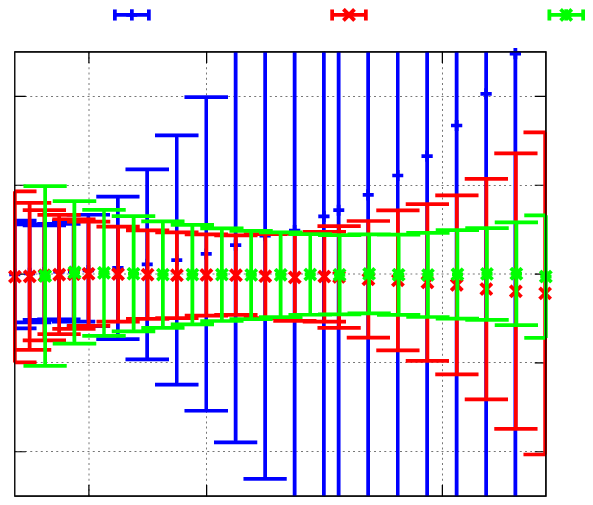}%
		\hfill
		\input{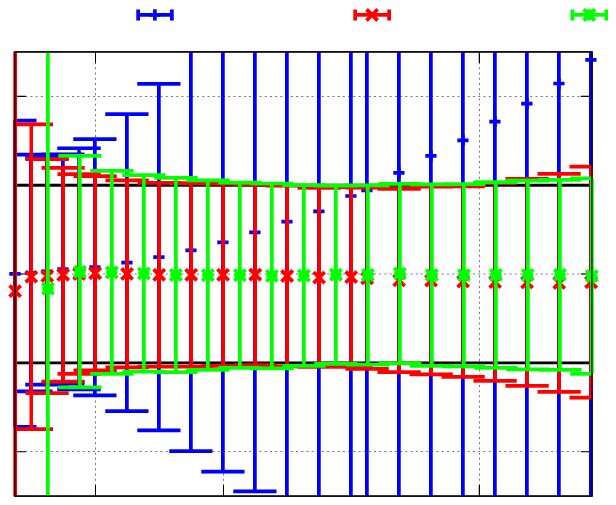}}%
	\caption{Elevation errors for different tracklet lengths using 5f-data.}
	\label{fig:lenele}
\end{figure*}

Concerning the elevation, shown in \Fig{fig:lenele}, the linear fit has the lowest and nearly constant absolute error for short tracklets. The quadratic fit reaches the level of the linear fit at approx. 40 s and afterwards maintains its accuracy whereas the linear one's is degrading rapidly. The relative error is reasonable for both fits up to approx. 40 s and thereafter only the quadratic fit maintains a consistent relative error. The fourth order fit becomes better for tracklets longer than approx. 120 s.

\subsubsection{Geocentric Inertial Positions}

\begin{figure*}[htbp]
	\centering
	{\input{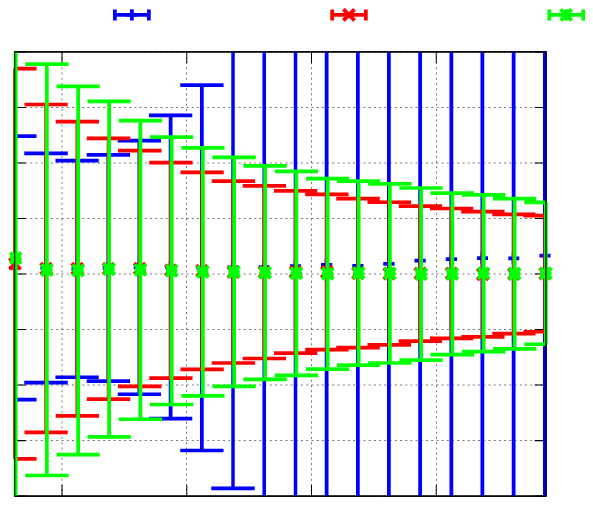}%
		\hfill
		\input{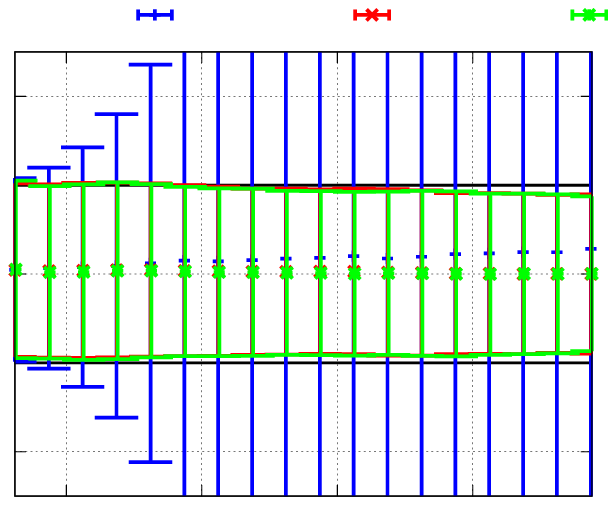}}%
	\caption{Position errors (X) for different tracklet lengths using 1f-data.}
	\label{fig:lenxpos}
\end{figure*}

\begin{figure*}[htbp]
	\centering
	{\input{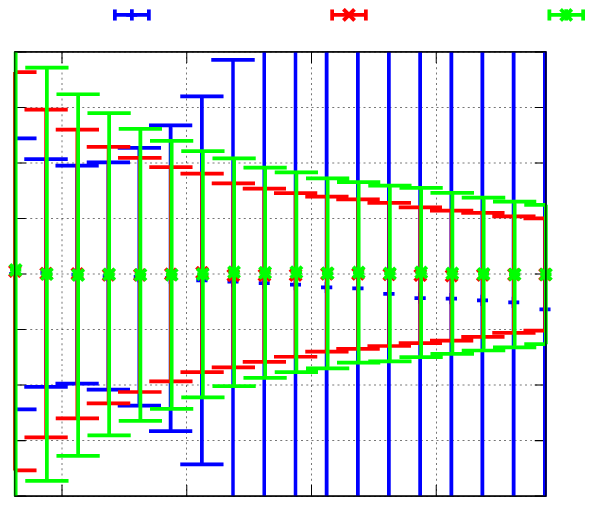}%
		\hfill
		\input{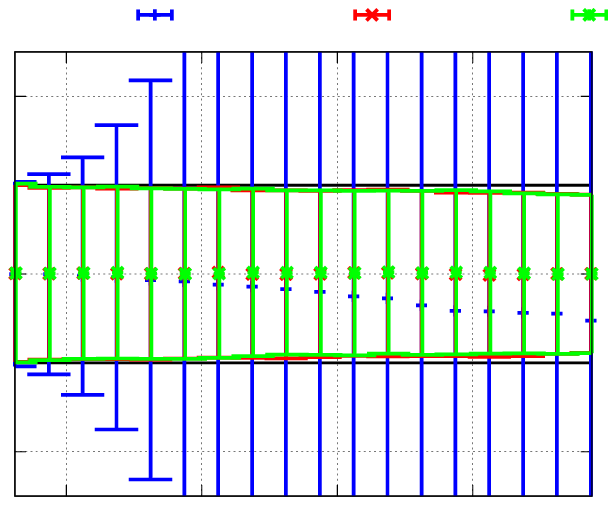}}%
	\caption{Position errors (Y) for different tracklet lengths using 1f-data.}
	\label{fig:lenypos}
\end{figure*}

\begin{figure*}[htbp]
	\centering
	{\input{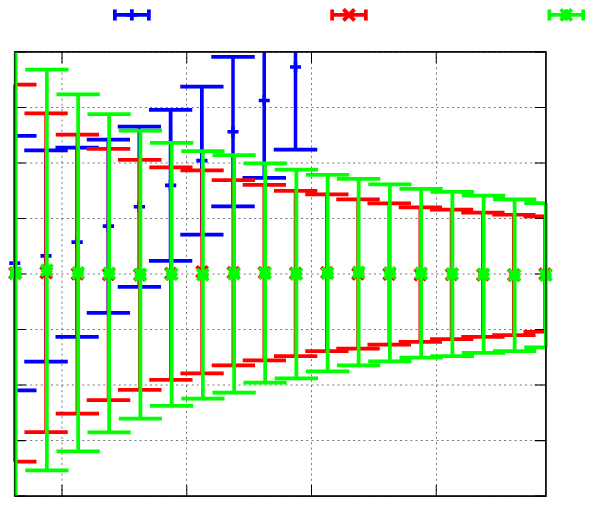}%
		\hfill
		\input{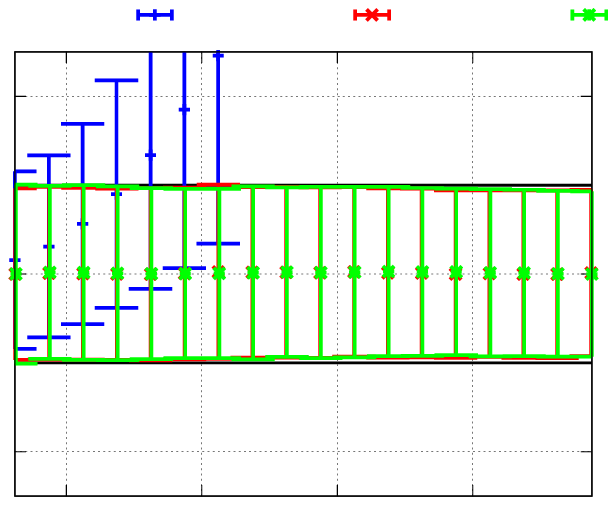}}%
	\caption{Position errors (Z) for different tracklet lengths using 1f-data.}
	\label{fig:lenzpos}
\end{figure*}

For the three geocentric inertial coordinate axes, the results are given in \Fig{fig:lenxpos}, \Fig{fig:lenypos} and \Fig{fig:lenzpos}. The X- and Y- coordinates are similar to each other and have a low error for the linear fit in the beginning, which increases with time after approx. 50 s. From there on the quadratic fit has the lowest error. Concerning the relative error, the one of the linear fit is increasing after 30 s while the quadratic one remains close to one within the checked range.

For the Z-position, the linear fit is also acceptable for short tracklets but it starts to build up a bias for longer tracklets. The linear fit cannot model the slight curvature of the coordinate over time, which leads to a consistent positive error in the Z-component due to the station location in the northern hemisphere, whereas for the X- and Y- coordinates there are both positive and negative errors which cancel out for the population mean. The quadratic fit is the best option for longer tracklets. As mentioned in \Sec{sec:acssys}, the main problem for this frame is the dependence on the orbit of the observed object, which makes it more difficult to have a consistent fitting approach.

\subsubsection{Topocentric AOS}

\begin{figure*}[htbp]
	\centering
	{\input{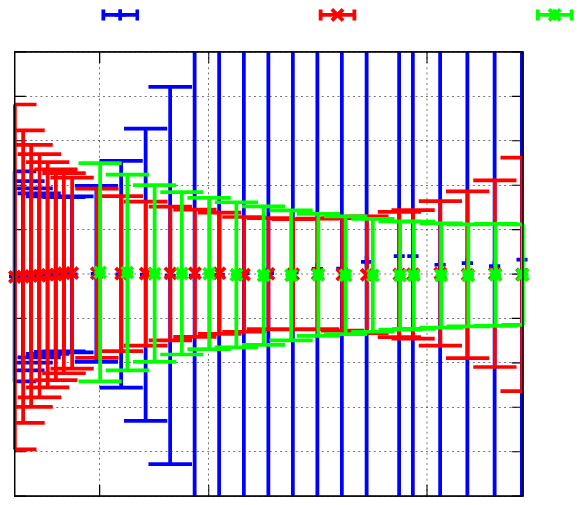}%
		\hfill
		\input{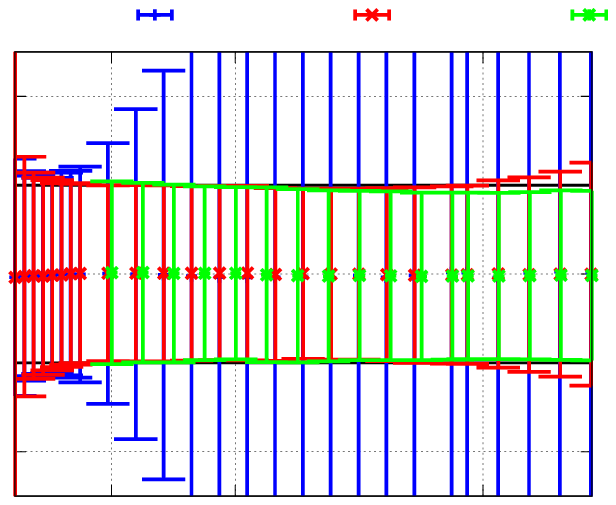}}%
	\caption{Azimuth errors (derived from topocentric AOS) for different tracklet lengths using 3f-data.}
	\label{fig:lenacstopip}
\end{figure*}

\begin{figure*}[htbp]
	\centering
	{\input{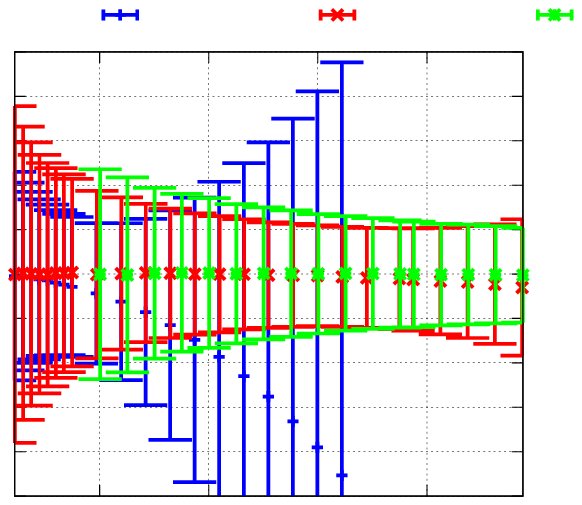}%
		\hfill
		\input{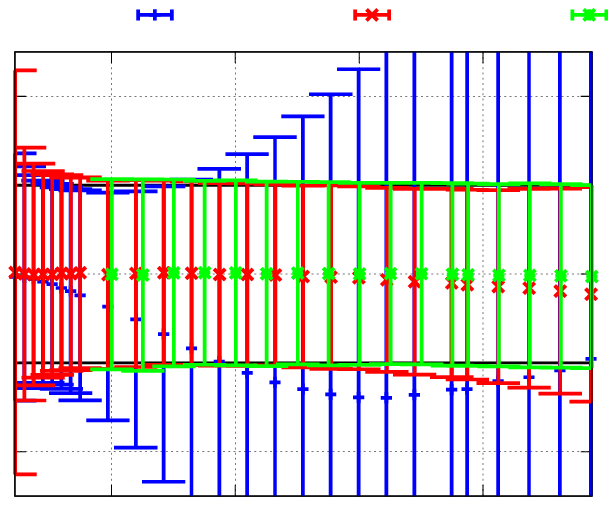}}%
	\caption{Elevation errors (derived from topocentric AOS) for different tracklet lengths using 3f-data.}
	\label{fig:lenacstopop}
\end{figure*}

As described in \Sec{sec:acssys}, the fitting of the attributable is done in the topocentric AOS but the angles are then transformed back to azimuth-elevation, which are used for the correlation and thus also used as a comparison to the ground truth in this section. The results are shown in \Fig{fig:lenacstopip} and \Fig{fig:lenacstopop}. For both observables, the linear fit is the best choice until approx. 30 s and afterwards the quadratic fit becomes better in terms of both absolute and relative error. For very long tracklets ($> 130$ s), the fitting should be extended to use a fourth order polynomial. Regarding the absolute error, it should be noted that especially for long tracklets an improvement compared to the direct azimuth fitting of at least a factor of five can be observed.

\begin{figure*}[htbp]
	\centering
	{\input{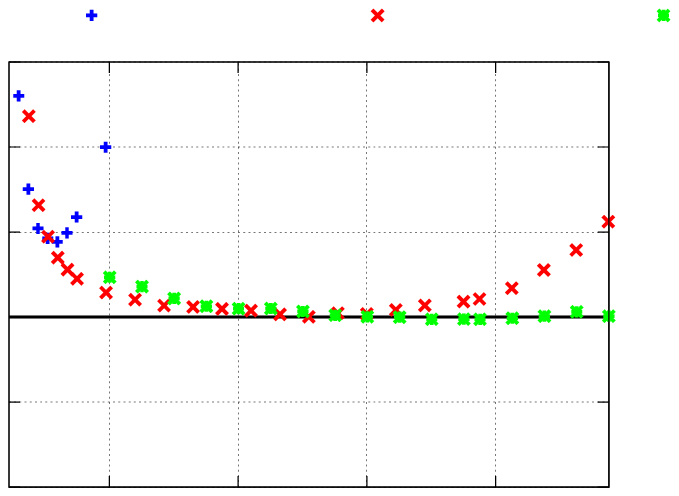}%
		\hfill
		\input{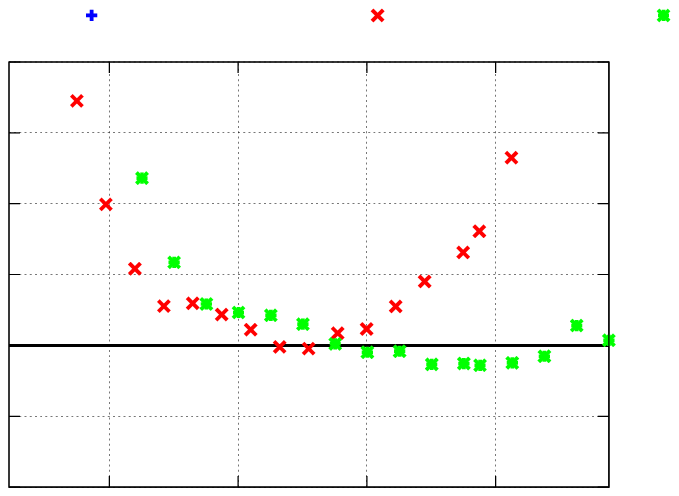}}%
	\caption{Mean and variance of Mahalanobis distances (derived from the topocentric AOS) of azimuth-elevation errors for different tracklet lengths using 3f-data. Theoretically expected values indicated by a bold line.}
	\label{fig:lenacstomaha}
\end{figure*}

Additionally, the back-transformation from AOS introduces a statistical correlation between the angular errors, which is checked via the Mahalanobis distance, see \Sec{sec:atteval}. The results for mean and variance are shown in \Fig{fig:lenacstomaha}. It can be seen that for the quadratic and fourth order fits using tracklets longer than 30 s, the values of the fitted population are close to the expected values indicated by the bold line, whereas for shorter tracklets and the linear fit this is not the case.

\subsubsection{Geocentric AOS}

\begin{figure*}[htbp]
	\centering
	{\input{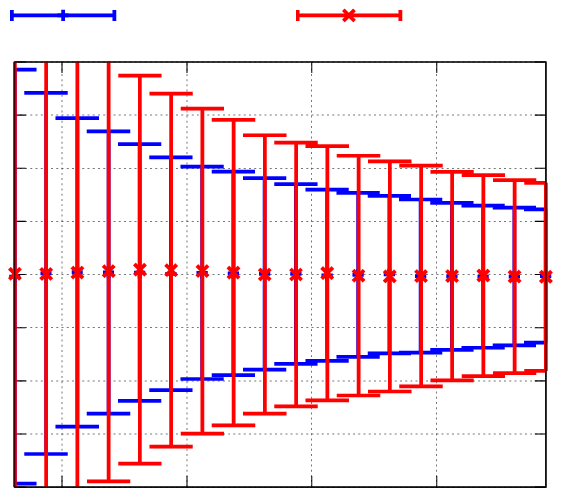}%
		\hfill
		\input{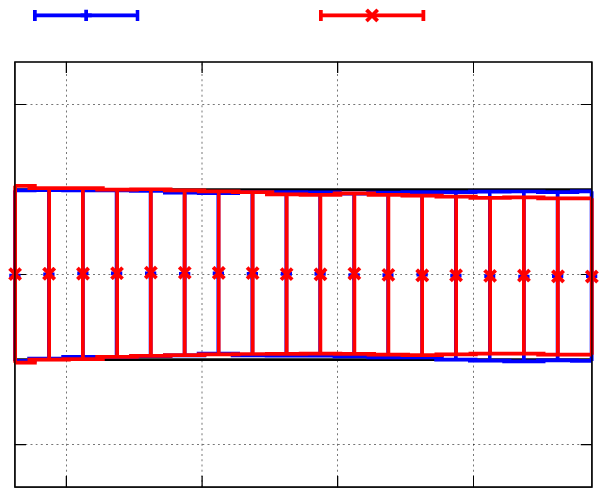}}%
	\caption{Position errors (X, derived from geocentric AOS)  for different tracklet lengths using 3f-data.}
	\label{fig:lenxposacsgeo}
\end{figure*}

\begin{figure*}[htbp]
	\centering
	{\input{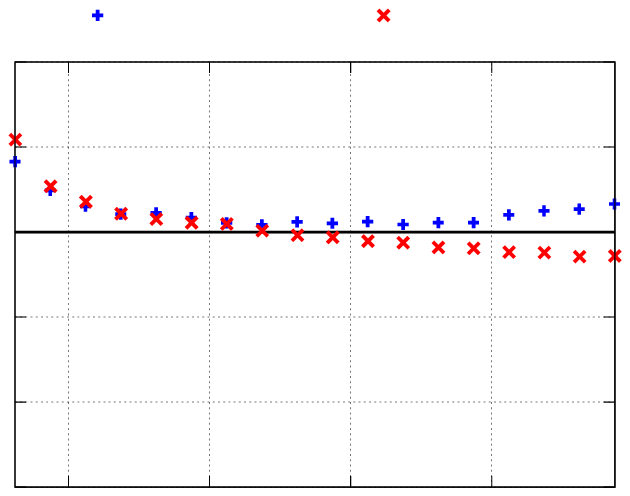}%
		\hfill
		\input{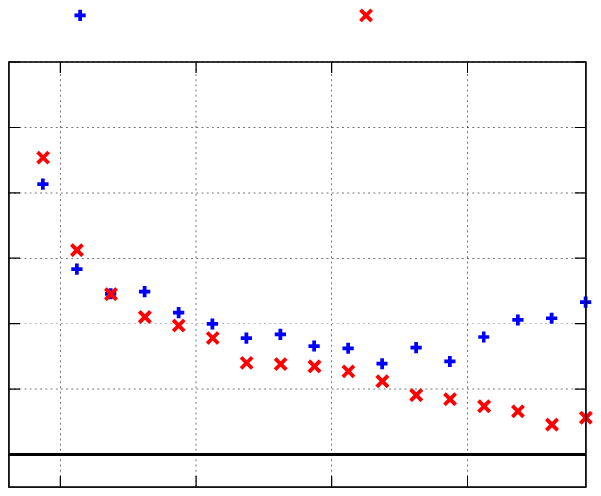}}%
	\caption{Mean and variance of Mahalanobis distances (derived from the geocentric AOS) of the XYZ-positions for different tracklet lengths using 3f-data. Theoretically expected values indicated by bold line.}
	\label{fig:lenacsgeomaha}
\end{figure*}

The geocentric AOS is transformed to geocentric inertial positions after the fit. Because the results for the three axes are similar, only the X-axes is shown in \Fig{fig:lenxposacsgeo}. It is visible that the linear fit is the best choice for all tracklet lengths. This suggests that the coordinates of this system are very stable over the pass. Although the absolute error is comparable to the direct fitting in XYZ, the higher consistency over the entire population is an improvement. Also here, the errors of the three positions become correlated which allows to test the distribution of the Mahalanobis distances, see \Fig{fig:lenacsgeomaha}. While the mean is well-approximated for the linear fit, the variance is approaching the theoretical value for longer tracklets but does not reach it. This could be due to the assumption of independent observables in the AOS during the fit. The effect of this will be investigated further in the \Sec{sec:corrs}.

\subsection{Order of the Fit}
\label{sec:order}

\begin{figure*}[htbp]
	\centering
	{\input{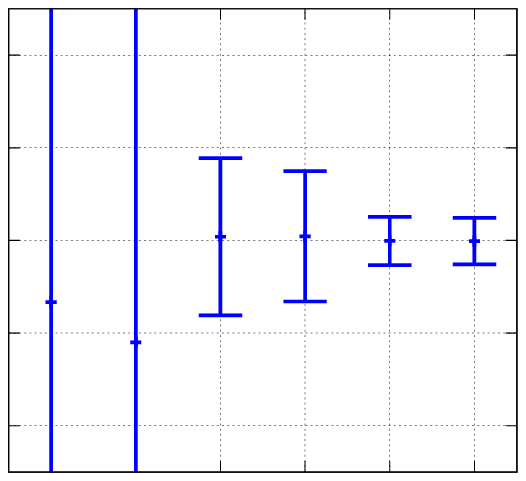}%
		\hfill
		\input{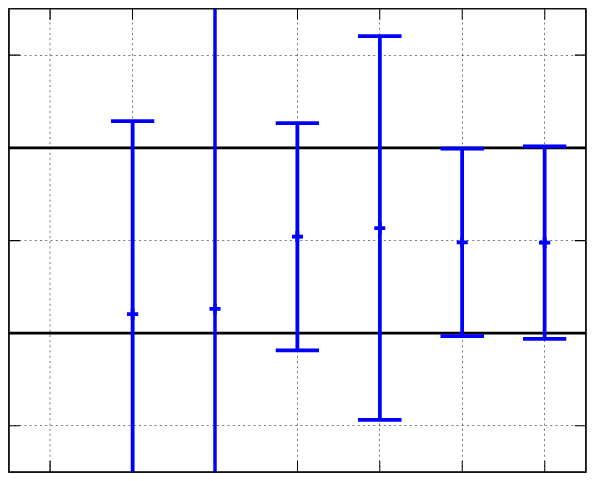}}%
	\caption{Range errors for different orders of the fitted polynomial using 3f-data (150 s).}
	\label{fig:ordrange150}
\end{figure*}

This subsection covers the potential use of even higher order polynomials for very long tracklets. The first example is the topocentric range, see \Fig{fig:ordrange150}, for a 150 s long tracklet. It can be seen that there is another step-like improvement of the absolute error if the order is increased to six. Additionally, it is visible that there is no steady improvement with the order of the fit, but the improvements are mainly dominated by adding even-numbered polynomial terms, which suggests that the measurements are line symmetrical around $\Delta t=0$ s.

\begin{figure*}[htbp]
	\centering
	{\input{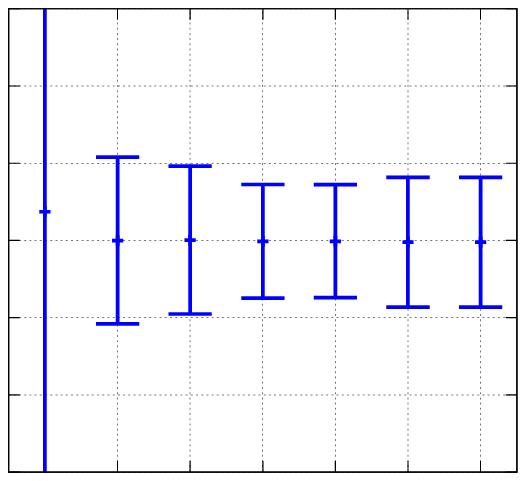}%
		\hfill
		\input{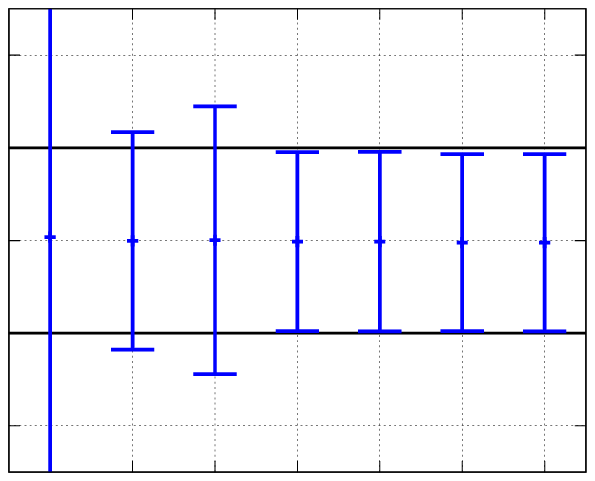}}%
	\caption{Azimuth errors (derived from topocentric AOS) for different orders of the fitted polynomial using 3f-data (150 s).}
	\label{fig:ordtopoazi150}
\end{figure*}

To contrast the previous example, \Fig{fig:ordtopoazi150} depicts the azimuth error for the topocentric AOS fit for the same tracklet length. There it is visible that going beyond the fourth order polynomial does not give an additional benefit. The analysis concerning different orders of the fitting has been performed for all observables. Higher orders have been added for the shown topocentric range and the directly-fitted azimuth angle. 

\subsection{Noise Level}
\label{sec:attnoise}

\begin{figure*}[htbp]
	\centering
	{\input{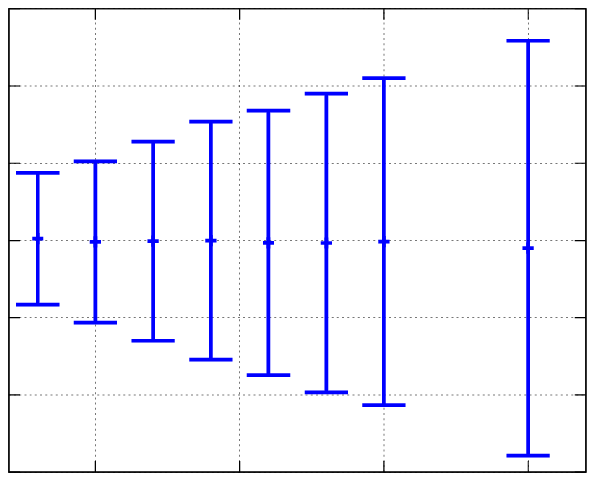}%
		\hfill
		\input{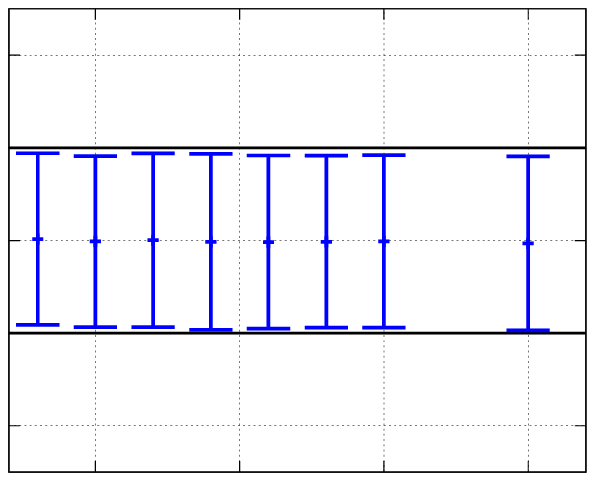}}%
	\caption{X-position errors for different noise levels using 3f-data (150 s) and a second order polynomial.}
	\label{fig:noisexyz150}
\end{figure*}

An integral part of the attributable fitting is the estimation of the uncertainties which relies on estimating the observable's measurement noise level from the fit's residuals. To show that this works for different levels of sensor noise, \Fig{fig:noisexyz150} compares the fitting results using a second order polynomial of the inertial X-coordinate for different noise levels by multiplying the values given in \Tab{tab:noise} with the factor on the abscissa. As one would expect, the absolute error is increasing with the noise level, whereas the relative error remains close to one suggesting that the estimation of the uncertainties is still correct.

\subsection{Range-rate via Range}
\label{sec:rrfit}

\begin{figure*}[htbp]
	\centering
	{\input{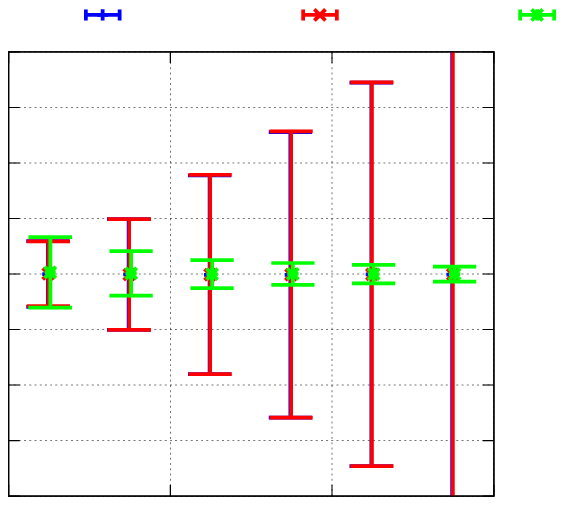}%
		\hfill
		\input{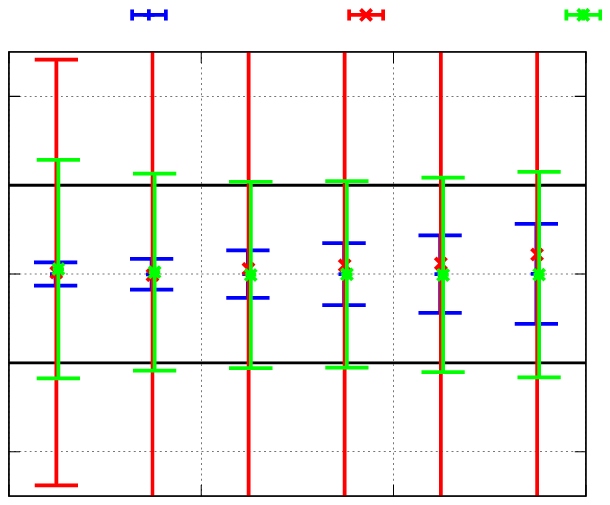}}%
	\caption{Range-rate errors estimated via the range derivative for different tracklet lengths using 1f data.}
	\label{fig:rrfit1f}
\end{figure*}

Another possibility is the assumption that the radar itself does not measure the range-rate via the Doppler shift, but it has to be calculated via the sequence of range measurements by using the first derivative in the parameter vector. The accuracy of the estimated range-rate using this approach is analysed for a linear, quadratic and fourth order fit. In \Fig{fig:rrfit1f}, one can see that the accuracy concerning the absolute error is comparable between the quadratic and linear fit, whereas that of the fourth order is much better and improving with more data. The order of magnitude of the error is smaller than for the pure range-rate fit previously shown in \Fig{fig:lenrate} which is due to a pessimistic assumption about the accuracy of the range-rate measurement. Concerning the relative error, it is obvious that the linear fit overestimates the uncertainty, whereas the quadratic fit underestimates the uncertainty, thus they have a relative error much lower and much larger than one, respectively. The fourth order polynomial leads to an acceptable relative error and could be used for the correlation. This would also mean that more data is necessary to perform the fourth order fit and get a good estimate of the range-rate, which may not be possible for very short tracklets. Because the range-rate is only used for the correlation decision and not in the orbit determination, the unrealistic error estimate may not degrade the correlation performance and only shifts the resulting Mahalanobis distances. This means that even with an unrealistic error of the range-rate, the correlation should still work although a different correlation threshold should be applied.

\subsection{Conclusion}

It was shown that the observables behave differently over the tracklet length and also compared to each other. The result of this analysis is a set of fitting rules for a group of tracklets with different lengths $t_T$. These rules are derived from the presented examples and are intended to keep the order of the polynomial as small as possible because especially for long tracklets with very sparse data, it may happen that there are not enough data points to use a high order polynomial. The resulting rules are presented in \Tab{tab:fitting}. From this summary, the geocentric AOS and the XYZ-positions seem to contain the most stable observables as they require the least different orders of the fit.

\begin{table}[htpb]
	\centering
	\caption{Rules of fitting depending on the tracklet length.}
	\begin{tabular}{|p{0.2\columnwidth} | p{0.12\columnwidth} | p{0.12\columnwidth} | p{0.12\columnwidth} | p{0.12\columnwidth} |}
		\hline
		Range- & $\leq$ 30 s & 30 s - 130 s &  \multicolumn{2}{|c|}{$\geq$ 130 s} \tabularnewline
		Rate & 1$^{\text{st}}$ order  & 2$^{\text{nd}}$ order  & \multicolumn{2}{|c|}{4$^{\text{th}}$ order}   \tabularnewline
		\hline
		\hline
		Range & $\leq$ 60 s & 60 s - 150 s & \multicolumn{2}{|c|}{ $\geq$ 150 s} \tabularnewline
		(topocentric) & 2$^{\text{nd}}$ order  & 4$^{\text{th}}$ order  & \multicolumn{2}{|c|}{ 6$^{\text{th}}$ order }  \tabularnewline
		\hline
		\hline
		Azimuth & $\leq$ 25 s & 25 s - 80 s & 80s - 150 s & $\geq$ 150 s \tabularnewline
		 & 1$^{\text{st}}$ order  & 2$^{\text{nd}}$ order  & 4$^{\text{th}}$ order & 6$^{\text{th}}$ order   \tabularnewline
		\hline
		Elevation & $\leq$ 40 s & 40 s - 120 s & \multicolumn{2}{|c|}{ $\geq$ 120 s} \tabularnewline
		& 1$^{\text{st}}$ order  & 2$^{\text{nd}}$ order  & \multicolumn{2}{|c|}{ 4$^{\text{th}}$ order }  \tabularnewline
		\hline
		\hline
		Inertial geocentric  &  \multicolumn{2}{c|}{$\leq$ 30 s} & \multicolumn{2}{c|}{$>$ 30 s} \tabularnewline
		Positions (X, Y, Z) & \multicolumn{2}{c|}{ 1$^{\text{st}}$ order} & \multicolumn{2}{c|}{ 2$^{\text{nd}}$ order }\tabularnewline
		\hline
		\hline
		Angles & $\leq$ 30 s & 30 s - 130 s & \multicolumn{2}{|c|}{ $\geq$ 130 s} \tabularnewline
		AOS topocentric & 1$^{\text{st}}$ order  & 2$^{\text{nd}}$ order  & \multicolumn{2}{|c|}{ 4$^{\text{th}}$ order }  \tabularnewline
		\hline
		\hline
		All observables  & \multicolumn{4}{c|}{all lengths} \tabularnewline
		AOS geocentric &\multicolumn{4}{c|}{ 1$^{\text{st}}$ order }\tabularnewline
		\hline
	\end{tabular}
	\label{tab:fitting}
\end{table}

\section{Results: Correlation}
\label{sec:corrs}
	
\subsection{Overview}

In the following, the influence of the attributables on the correlation performance is analysed which is the overall purpose of the fitting process. In these experiments we used a subset of 1000 tracklets to reduce the workload of the correlation. Two quantities are used to assess the quality of the correlation. One plot compares the correlation performance in terms of the percentage of detected true positives (TP), calculated as the number of identified true correlations divided by the total number of true correlations in the sample, together with the percentage of false positives (FP), obtained by dividing the number of identified false correlations by the total number of identified correlations. The second plot depicts the standard deviation $\sigma$ of the orbital elements semi-major axis, eccentricity, inclination and Right Ascension of the Ascending Node (RAAN) $\Omega$, derived by analysing the differences between the estimated value and the true value given by the orbit which was used to simulate the noisy measurements. These standard deviations are cleared from 3-$\sigma$ outliers as it was already explained for the attributables. It is important to note that these orbits are now the result of the combination of two attributables and thus the lengths of the tracklets are only contributing indirectly. The initial orbit determination and correlation is performed according to the method in \cite{reihs2020method}. This method consists of a J$_2$-perturbed initial orbit determination from two geocentric inertial positions which are derived from the attributables. The range-rate which would be observed for this initial orbit is compared to the measured one from the attributable to calculate the Mahalanobis distance $M_\text{d}$. The following tests assume a correlation threshold of $M_\text{d}=3$, which would theoretically include approx. 99\% of true positives according to the $\chi$-distribution.

% 1 ................................................................................
\subsection{General Comparison}

\begin{figure}[htbp]
\centering
\input{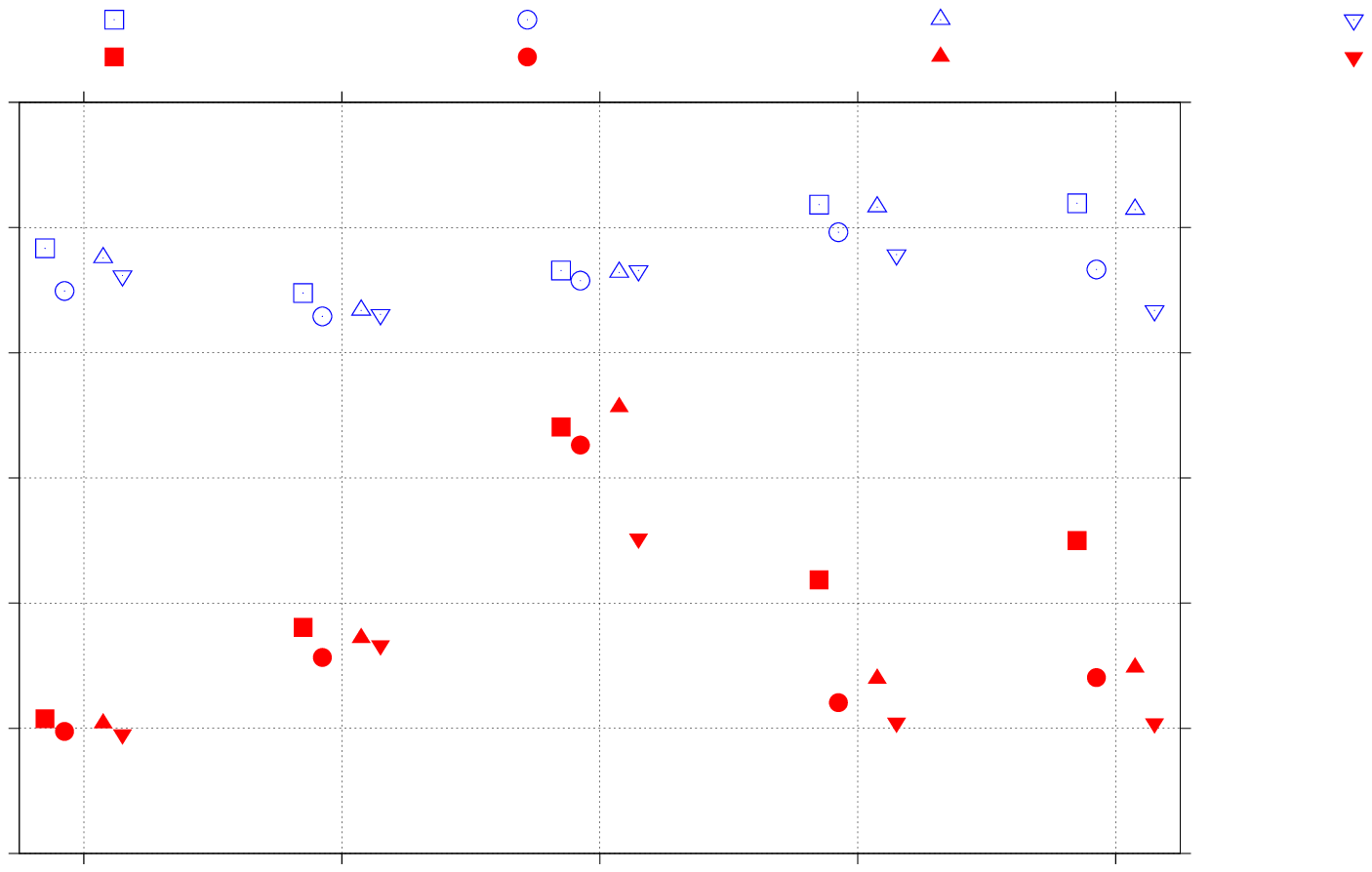}
\caption{Comparison of percentages of true and false positives for different coordinate systems and test cases.}
\label{fig:compgen}
\end{figure}

The first comparison uses five different combinations of tracklet length and detection frequency to give a first impression of the performance of the different coordinate systems. The results comparing the percentages of true and false positives are shown in \Fig{fig:compgen}. Starting from a short and dense tracklet, one can see that the differences between the systems are increasing with the length of the tracklets. Concerning the true positives, the results are similar for the first three cases and start to diverge after that. The geocentric AOS has the lowest percentage of true positives for the last two cases, but also the lowest values for the false positives. This might be due to a shift of the distribution of Mahalanobis distances towards higher values, which could be explained by the larger than expected variance of the fitted position errors' Mahalanobis distance shown previously in \Fig{fig:lenacsgeomaha}.

\begin{figure*}[htpb]
	\centering
	\subcaptionbox{10s 1f data}
	{\input{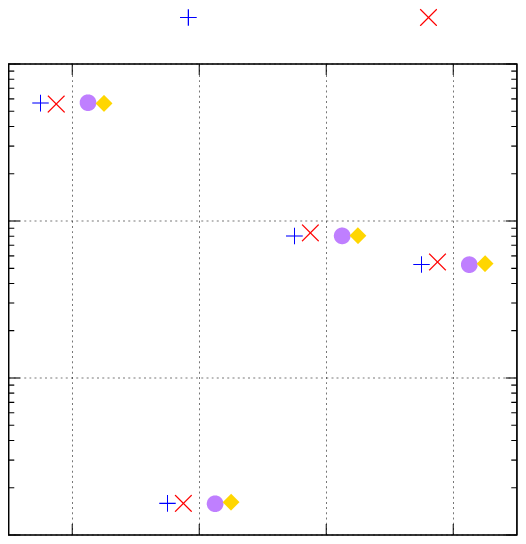}}%
	\hfill
	\subcaptionbox{30s 5f data}
	{\input{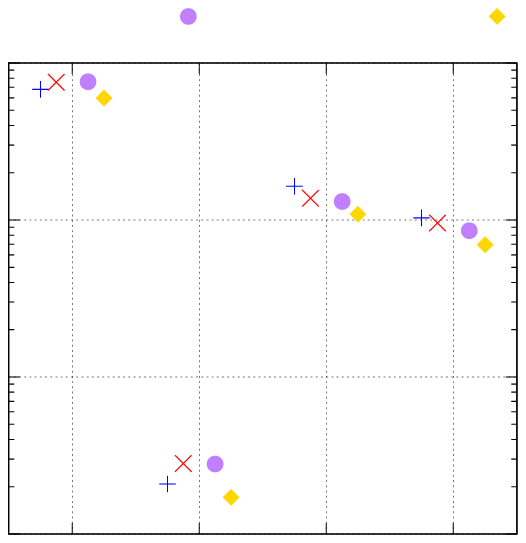}}%
	\vfill
	\subcaptionbox{100s 5f data}
	{\input{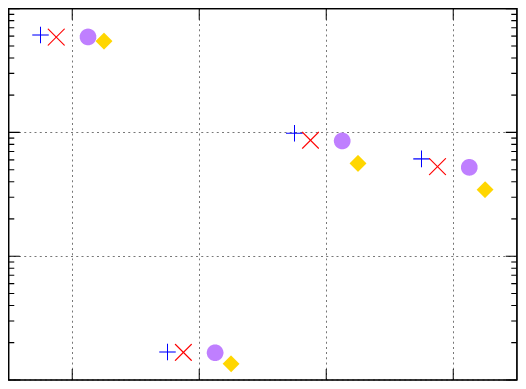}}%
	\hfill
	\subcaptionbox{180s 5f data}
	{\input{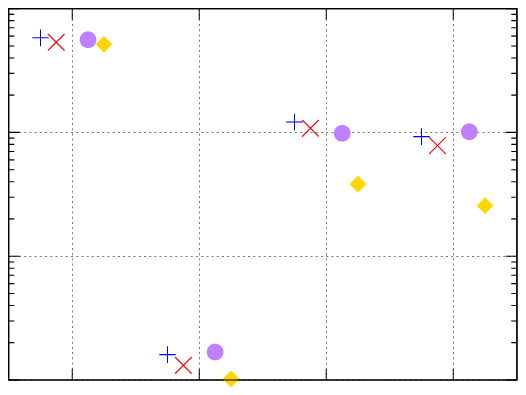}}%	
	\caption{Orbit errors of true positives for different coordinate systems and test cases (respective units are given on the abscissa).}
	\label{fig:yerrgen}
\end{figure*}

The orbital accuracies for four of the previous examples are compared in \Fig{fig:yerrgen}. The largest differences between the coordinate systems is in the estimation of the orbital plane, namely the inclination and RAAN. For these elements, the geocentric AOS is clearly the best system, while the azimuth-elevation system is the worst. The differences between them increase with the tracklet length.

% 2 ................................................................................
\subsection{Influence of Tracklet Length}
\label{sec:sucorlen}

\begin{figure*}[htbp]
	\centering
	\input{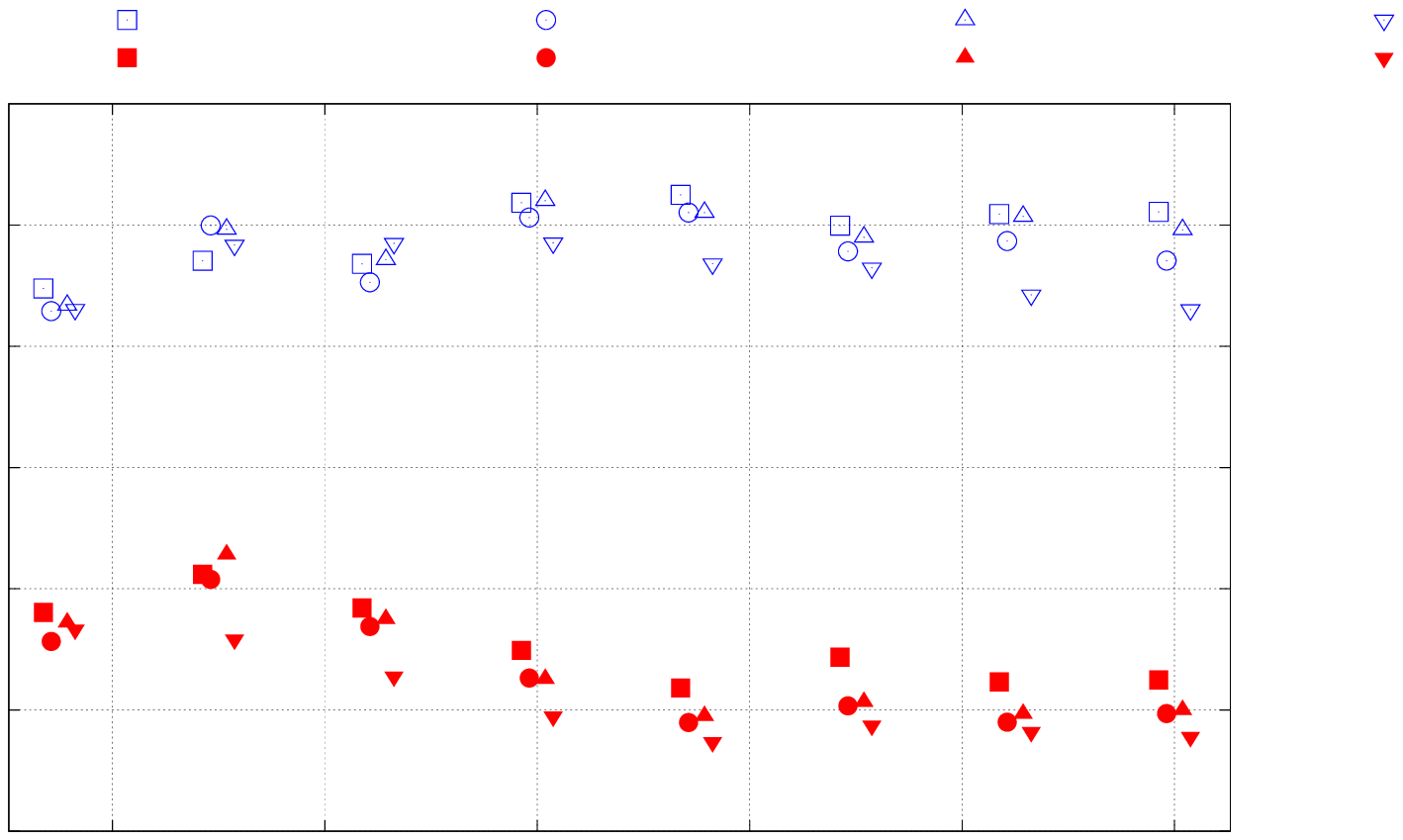}%
	\caption{Percentages of true and false positives using 3f-data for different coordinate systems over a varying tracklet length. Data points are in 15 s intervals from 15 s to 120 s and the different systems are horizontally separated to improve readability.}
	\label{fig:lenerror1f}
\end{figure*}

After the results in the previous section already indicated a dependence on the tracklet length, this effect is investigated in the following, see \Fig{fig:lenerror1f}. Using the data with 3 seconds between detections, tracklet lengths of up to two minutes are compared. For tracklets longer than 45 s, the percentage of true positives is around 95\% with a slight trend towards lower values, most strikingly for the geocentric AOS. Again, this is probably due to the approximated statistical correlation coefficient in the covariance matrix and this effect becomes larger when the estimated uncertainties are decreasing due to the increased tracklet length. Another possible explanation is discussed in \Sec{sec:bias}. The geocentric AOS also has the lowest percentage of false positives, which is around 5\% for all systems. The azimuth-elevation system has most false positives.

\begin{figure*}[htpb]
	\centering
	\subcaptionbox{Semi-major axis}
	{\input{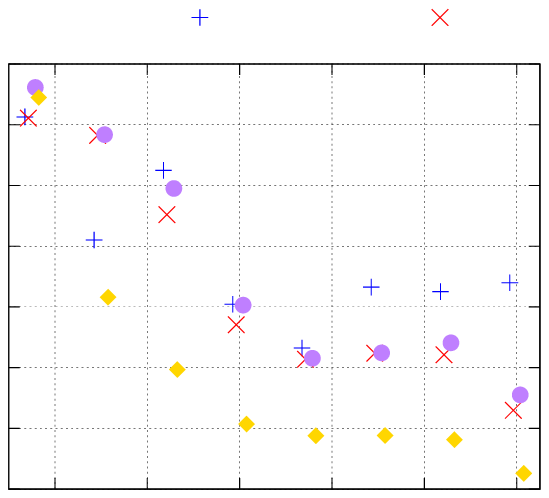}}%
	\hfill
	\subcaptionbox{Eccentricity}
	{\input{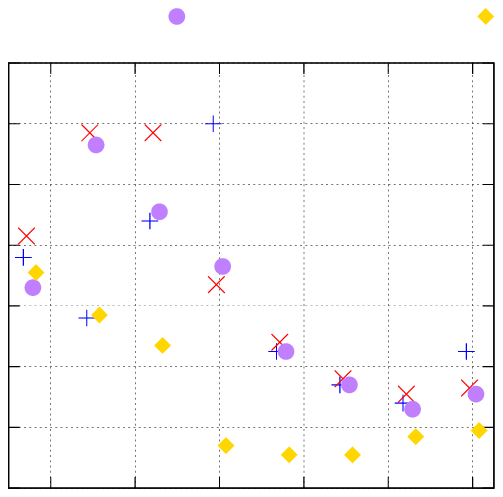}}%
	\vfill
	\subcaptionbox{Inclination}
	{\input{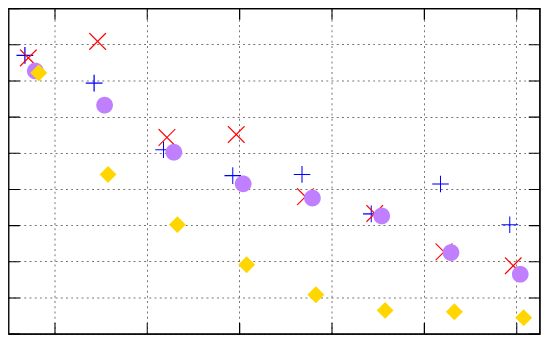}}%
	\hfill
	\subcaptionbox{RAAN}
	{\input{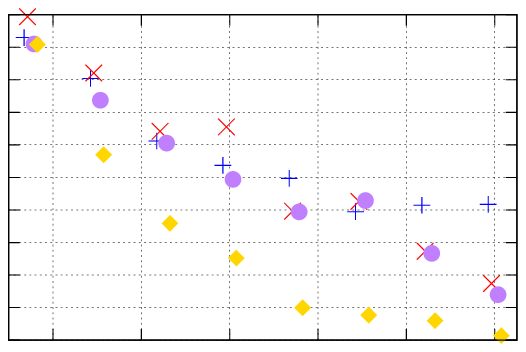}}%	
	\caption{Orbit errors of true positive for different coordinate systems over the tracklet length.}
	\label{fig:yerrlengen}
\end{figure*}

The orbital accuracy also depends on the tracklet length as shown in \Fig{fig:yerrlengen}. It should be noted first that all elements and systems have a general trend towards decreased errors, which indicates that the increased amount of information in a longer tracklet is transported via the attributable until the orbit level. Comparing the different systems, the geocentric AOS has the highest level of accuracy again. Especially for long tracklets, the azimuth-elevation system is much worse than the rest. Also here it is visible that the differences increase with the tracklet length. For very short tracklets (15 s), there are no significant differences between the systems. Thus it can be concluded that the use of derived observables in different coordinates is mainly beneficial for tracklets of at least 20 s.

% 3 ................................................................................
\subsection{Influence of Detection Frequency}

\begin{figure*}[htbp]
	\centering
	\input{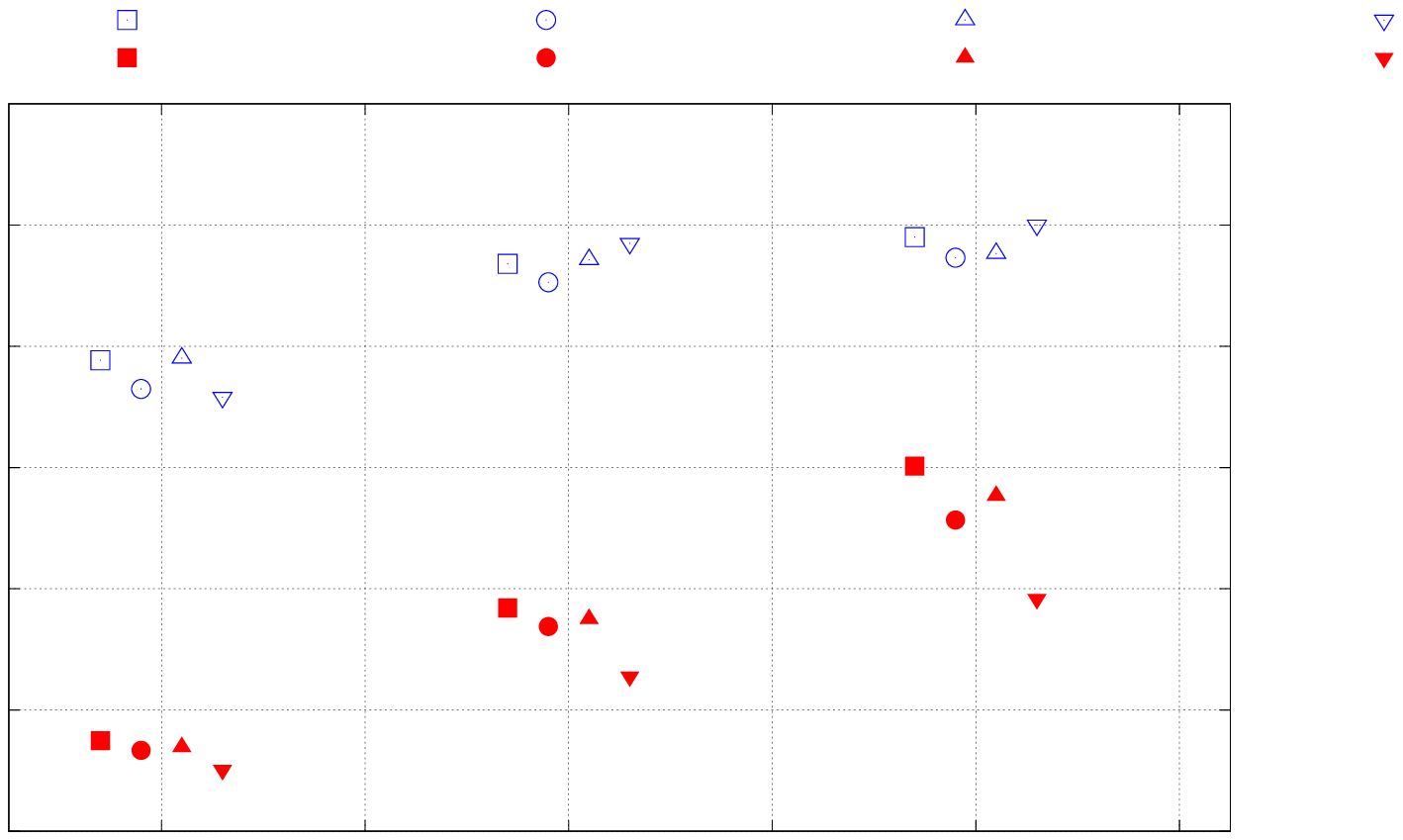}%
	\caption{Percentages of true and false positives using 45 s tracklets for different coordinate systems and detection frequencies (1, 3, 5).}
	\label{fig:corfreq}
\end{figure*}

Another possible variation of the input data is the frequency of detections, which is mainly dependent on the size of the FoR and the scanning strategy. The effect of this parameter is assessed based on a specific tracklet length with different frequencies and thus different numbers of detections. \Fig{fig:corfreq} compares the three detection frequencies for a tracklet length of 45 s. From the attributable fitting, it was concluded that less data points lead to an increase of the observables' uncertainty. This increased uncertainty leads to more false positive correlations. The same effect also leads to a slight increase of true positives.  Although all three elements used for the initial orbit have an increased uncertainty, single elements can still be relatively good. If e.g. one of the angles, which are the dominant source of error, is estimated well with a large uncertainty, this improves the orbit determination while still having the large uncertainty. This effect can lead to the slight increase of true positives by including those combinations which are just larger than the threshold for smaller uncertainties.

\begin{figure*}[htbp]
\centering
	\input{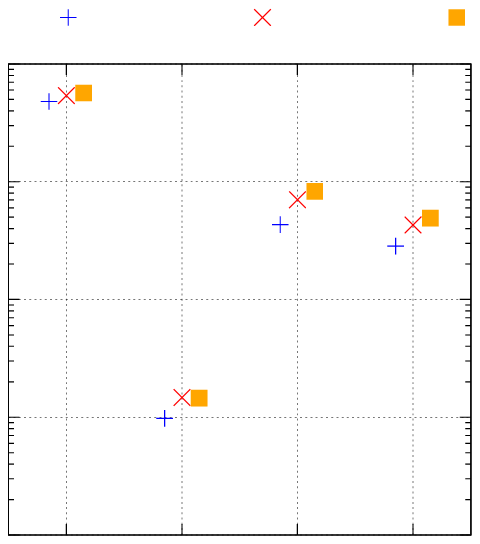}%
\caption{Orbit errors of true positives using 45 s tracklets in the geocentric AOS over different detection frequencies.}
\label{fig:corfreqorb}
\end{figure*}

In \Fig{fig:corfreqorb}, the orbit errors are compared only for the geocentric AOS, as the other results are similar. As expected, the increased uncertainty due to less data leads to larger errors in the orbital elements.

% 4 ................................................................................
\subsection{Noise Levels}
\label{sec:cornoise}
\begin{figure*}[htbp]
	\centering
	{\input{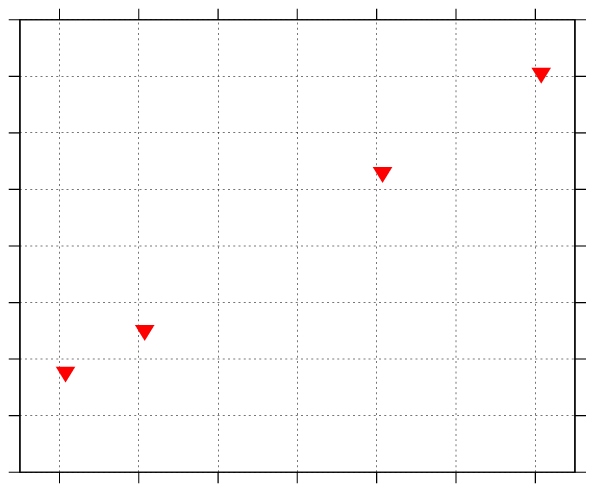}%
		\hfill
		\input{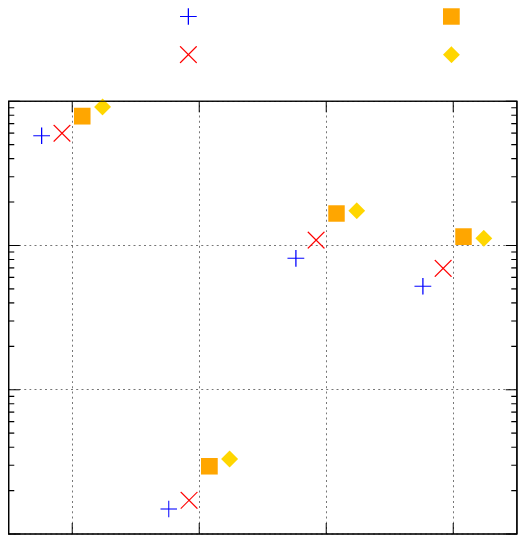}}%
	\caption{Percentages of false positives and orbit errors for different noise levels in the geocentric AOS (30 s, 5f).}
	\label{fig:cornoise}
\end{figure*}

In order to show the effect of the increased noise, which was already considered for the attributables in \Sec{sec:attnoise}, only the example of the geocentric AOS is shown in \Fig{fig:cornoise} using a sparse 30 s tracklet with 5f-data. The results for the other systems are similar. The results include the percentage of false positives, which shows a nearly linear trend over the increased noise, and the errors of the orbit determination which are also increasing with the noise level. Both effects are consistent with intuitive expectations.

% 5 ................................................................................
\subsection{Range-Rate via Range}

Although not specifically shown here, the correlation experiments were also performed for examples with a range-rate via the range derivative as discussed in \Sec{sec:rrfit}. Using the fourth order range fit for tracklets with a length of 45 s (3f-data), the results are similar to those presented in \Fig{fig:lenerror1f}. The main difference is the lower value for the percentage of false positives due to the smaller uncertainty of the range-rate. Thus, the correlation is also possible if the range-rate is estimated via the range. For the remaining experiments, the range-rate is assumed to be measured again.

\section{Survey Campaign}
\label{sec:survey}

\subsection{Simulation Parameters}

For the final experiments of this paper, the correlation results for the complete surveillance scenario introduced in \Sec{sec:simsce} are evaluated. The parameters were given in \Tab{tab:noise} with the resulting distribution of tracklet lengths given in \Fig{fig:dwell} combined with a cut-off at \mbox{180 s} to avoid very long tracklets as outliers. Considering the previous results in this paper, it is decided that the geocentric AOS is the best coordinate system for the correlation and thus the survey results mainly use these coordinates. In addition to the already used measurement frequencies of one (1f), three (3f) and five (5f) seconds between the detections, there is one additional test with 8 s between detections to further test the robustness of the correlation. This yields the total numbers of pairs and true correlations given in \Tab{tab:corrs}. Tracklets which do not have enough data points to fit the polynomial of the required order are removed (for a linear fit a minimum of three detections), which explains the reduced number of tracklets for longer detection intervals. As a comparison, the other three coordinate systems are tested with the 3f-data only. In order to reduce the number of false negative correlations, which describes the not identified true correlations, the threshold for the following tests is set to $M_d = 5$. This is necessary due to the shift of Mahalanobis distances to larger values, which is due to a specific characteristic of the J$_2$-perturbed orbit determination. This is discussed in \Sec{sec:bias}. Also the already mentioned bias due to the correlation of the errors before the fit, see \Fig{fig:lenacsgeomaha}, is introducing an offset for the geocentric AOS.

\begin{table}[htpb]
	\centering
	\caption{Total combinations and true correlations for the geocentric AOS using different detection frequencies.}
	\begin{tabular}{|p{0.1\columnwidth} | p{0.15\columnwidth} | p{0.2\columnwidth} | p{0.2\columnwidth}|}
		\hline
		Scenario & Tracklets & Total Combinations & True Correlations \\
		\hline
		\hline
		1f & \centering 4 772 & \centering 11 383 606 & \centering 717 \tabularnewline
		\hline
		3f & \centering 4 598 & \centering 10 568 503 & \centering 683 \tabularnewline
		\hline
		5f & \centering 4 384 &\centering $\ $ 9 607 536 & \centering 636 \tabularnewline
		\hline
		8f & \centering 4 055 &\centering $\ $ 8 219 485 & \centering 568 \tabularnewline
		\hline
	\end{tabular}
	\label{tab:corrs}
\end{table}

\subsection{Additional Filters}

As it was shown in \Tab{tab:corrs}, the number of true correlations over 24 hours is very low compared to the total number of pairs which have to be checked. This is the main difference to the previously presented tests, when the share of true correlations was much higher. This low number of correlations combined with many objects on similar orbits, especially sun-synchronous polar orbits, leads to a very high number of false positives after the standard attributable-based correlation step. To remove these false positives, two additional filters are added after the attributable correlation to reject false positives.

\begin{figure}[htbp]
\centering
\input{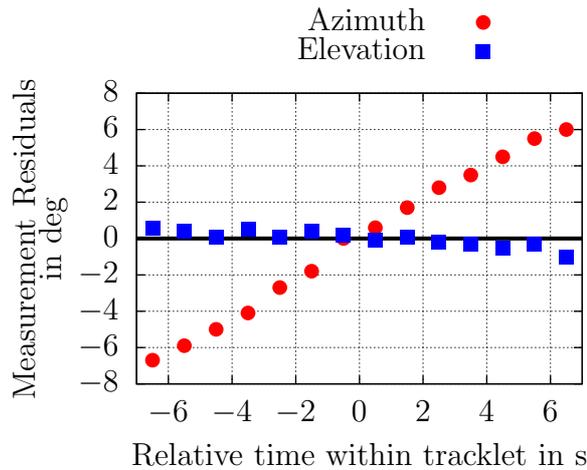}
\caption{Example of the residuals in azimuth and elevation after the initial orbit determination based on attributables for a single tracklet.}
\label{fig:aziexample}
\end{figure}

Firstly, it has to be considered that the attributable given in \Equ{eq:attri}, does not consider the direction of motion, i.e. the derivatives of the angles. This is not done, because the angular measurements have high uncertainties and thus the estimation of their derivatives has a large uncertainty which may not be reflected properly in the fit. If two tracklets are correlated via their attributables, the underlying full tracklets can be used to calculate the residuals of the single observations based on the resulting orbit of the correlation. An example is given in \Fig{fig:aziexample} showing the residuals of azimuth and elevation for a false positive correlation. The linear trend in the azimuth direction is clearly visible which indicates that the orbit and the tracklet are not matching. To identify these pairs reliably, a Student's t-test \cite{bhattacharyya1977statistical} checks if there is a linear trend in the residuals. If such a trend is detected, this correlation is removed, see \cite{fujimoto2014association} for further details. While the t-test works well for long tracklets, it may not be possible to identify linear trends for shorter tracklets.

\begin{figure}[htbp]
	\centering
	\input{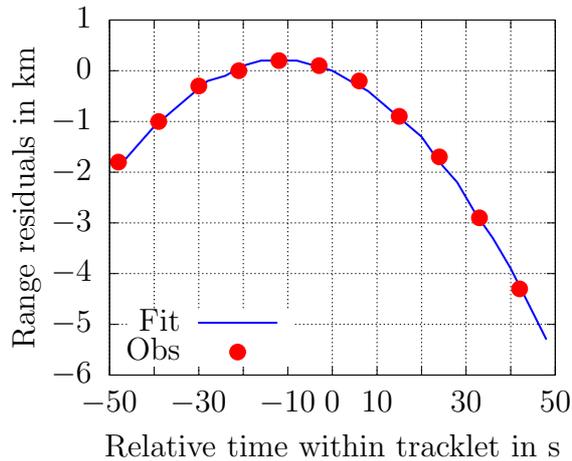}
	\caption{Range residuals of the initial orbit (number of measurement points reduced for better readability).}
	\label{fig:pprho}
\end{figure}

A comparable test is done using the range measurements. \Fig{fig:pprho} depicts the range residuals of a tracklet against the orbit which was calculated from the attributable correlation. In this case, it is a true positive correlation but with a wrong estimation of the number of revolutions between the detections which leads especially to a wrong estimate for both the semi-major axis and the eccentricity. This causes that the range residuals over the entire tracklet become biased, but the zero residual at the reference time ($\Delta t=0$ s) confirms that at least the attributable is matching the orbit. Ideally for a true correlation with a correct orbit, these residuals should be close to a random normal distribution consistent with the sensor noise. In order to extract this information and test if the residuals are biased, a second-order polynomial is fitted to the range residuals:
\begin{equation}
\Delta \rho = \frac{1}{2} \cdot c_2 \cdot \Delta t^2 + c_1 \cdot \Delta t + c_0 \ .
\end{equation}
The absolute value of the curvature $|c_2|$, here defined as the second derivative, is used as a further parameter for the correlation decision. If this curvature is larger than the threshold value $c_{2, \textrm{thresh}}=0.002$, the correlation is discarded. This is mainly useful if there are several days between the tracklets and the semi-major axes of adjacent solutions are close to each other. For the 24-hours survey analysed here, this is of minor importance but reported for completeness.

For correlations which pass both thresholds, for t-test and curvature, these information can even be combined to get an additional information. The idea is that the result $p_t$ from the t-test should be as large as possible while the absolute value of the curvature $|c_2|$ should be as small as possible. This can be combined into a joint score $S$ for both tracklets (T1 and T2), which becomes larger if the tracklets match the orbit better:
\begin{equation}
S = \frac{p_{t,\text{T1}} + p_{t,\text{T2}}}{|c_{2,\text{T1}}| + |c_{2,\text{T2}}|} \ .
\end{equation}
The value of this score has no general objective meaning but it can be used to choose between solutions with different numbers of revolutions for the same pair of tracklets. Practically, this means that it is no longer only the solution with the smallest Mahalanobis distance, which is considered, but all solutions which are smaller than the threshold. Then, the score $S$ is calculated for each solution and the one with the largest score is chosen.

A final stage of a least squares orbit determination using both tracklets is performed. The correlation is excluded, if this orbit calculation fails or the residuals of the measurements are too large. The mean and standard deviation of the residuals is used as a threshold. It is set for the angular observables at $\mu_{T,LS,A}= 0.11^\circ \  (0.17^\circ)$ and $\sigma_{T,LS,A}= 0.3^\circ \ (0.4^\circ)$. The values in parentheses are used for the experiments using 5f- and 8f-data to consider the increased uncertainty due to less data in the tracklet. The thresholds for the ranges are set at $\mu_{T,LS,\rho}= 10$ m and $\sigma_{T,LS,\rho}= 40$ m. These values are chosen slightly higher than the known performance of the radar, see \Tab{tab:noise}, to account for the uncertainty due to the relatively small amount of information from two passes. Only correlations which pass all three steps are accepted.

\subsection{Results}

\subsubsection{Overview}

\begin{figure*}[htpb]
	\centering
	\subcaptionbox{1f-data}
	{\input{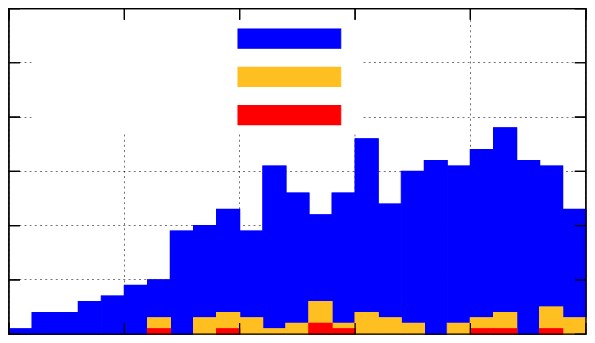}}%
	\hfill
	\subcaptionbox{3f-data}
	{\input{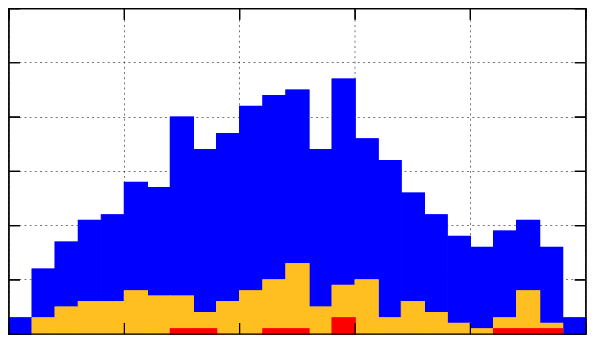}}%
	\vfill
	\subcaptionbox{5f-data}
	{\input{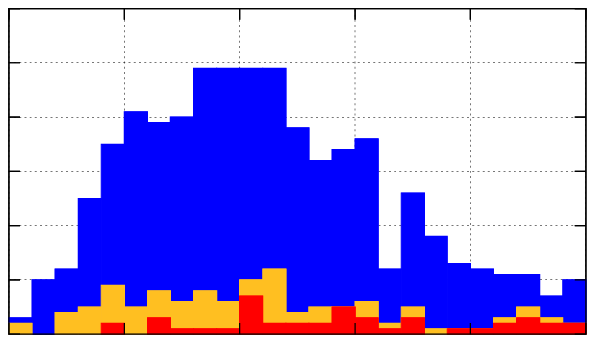}}%
	\hfill
	\subcaptionbox{8f-data}
	{\input{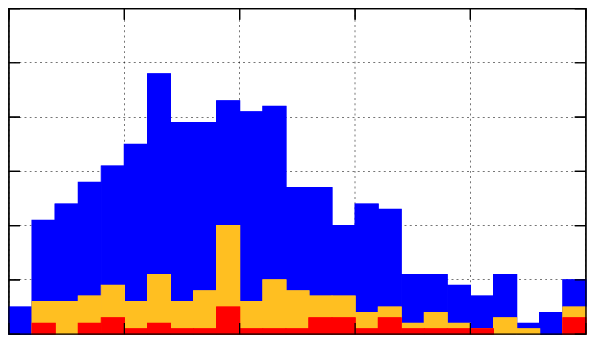}}%	
	\caption{Histograms of correlation results for the surveys with different detection intervals.}
	\label{fig:surveys}
\end{figure*}

The processing of the experiments takes approx. 24 - 27 hours per simulation on a computer with four Intel Core i5-3470 CPU (3.20 GHz) running three processes in parallel. Because parallelisation of the pairwise correlation problem is simple, more parallel processes on a capable computer could be used to reduce the processing time further, which should be of no concern for an operational system. The results of the survey experiments are presented in \Fig{fig:surveys}. In the plot, true positive (TP) refers to the confirmed correlations after the least squares and false negative (FN) refers to the true correlations, known from the simulation, which were found but did not lead to a converged least squares with sufficiently small residuals. All results show a distinct peak of true positives, which shifts towards lower Mahalanobis distances with an increasing observation interval. This will be explained in the following subsection. All experiments also have a share of false positives and non-converged least squares, which also increases with more time between the detections.

\Tab{tab:res} summarises all results for the different experiments. Two main effects are visible. The first is, that the share of detected true positives increases with the detection interval, because the tail of the distribution which is cut at the threshold becomes less due to the mentioned shift of the Mahalanobis distances. For the 1f-data, an increase in the $M_\text{d}$-threshold would also lead to more TP, but at a higher computational cost because more FP would require a further test with the least squares orbit determination. Opposed to that, the share of converged least squares orbit determinations decreases with an increased detection interval, because less data is available in the tracklets. To counteract this, the acceptance threshold of the least squares has been increased for the 5f- and 8f-data, which also leads to a significant increase in false positives. Without the increased threshold, the FP-level for the 8f-data would also be around 2 \% but with only 50 \% converged TP. 

As a comparison, \Tab{tab:res} also shows the results for the other three coordinate systems using the 3f-data. Compared to the geocentric AOS, it is visible that the percentage of true positives is larger for the other systems, especially directly after the attributable correlation. This suggests that the shift in the Mahalanobis distances is only partially due to the bias explained in the next section, but also caused by the missing consideration of the error correlation as shown in \Sec{sec:attris}. Increasing the correlation threshold to $M_d=7.5$ would let the geocentric AOS also reach the 99.8\% of true positive correlations like the other systems at a false positive rate of still only approx. 95 \% before the least squares. The number of false negative correlations due to a non-converging least squares is similar for all coordinate systems. Thus the main difference between the coordinate systems is the efficiency with regard to the minimisation of false positives to reduce the computational effort, but the resulting number of confirmed correlations is similar for all the systems. This may also be due to the definition of the FoR which has a maximum elevation of $70^\circ$. If a radar's FoR includes $el=90^\circ$, thus the singularity in the azimuth-elevation system, tracklets passing close to this singularity might lead to more differences between the results of different coordinate systems.

\begin{table}[htpb]
	\centering
	\caption{Results of the correlation of survey campaigns. (*) indicates a higher acceptance threshold for the least squares, see text.}
	\begin{tabular}{|p{0.15\columnwidth} | p{0.1\columnwidth} || p{0.1\columnwidth} | p{0.1\columnwidth}|| p{0.12\columnwidth} | p{0.1\columnwidth}|}
		\hline
		Coordinates & Freq. & \multicolumn{2}{|c||}{After Attr.}  & \multicolumn{2}{|c|}{After LSQ} \\
		 & & TP & FP & TP & FP\\
		\hline
		\hline
		geo. AOS & \centering 1f & 73.9 \% &  81.4 \% & 67.8 \% &  1.6 \%  \\
		\hline
		geo. AOS & \centering 3f & 95.5 \% &  90.6 \% & 76.7 \% &  1.9 \%  \\
		\hline
		geo. AOS & \centering 5f & 97.0 \% &  93.8 \% & 85.2 \% (*)&  7.2 \% (*) \\
		\hline
		geo. AOS & \centering 8f & 96.0 \% &  96.1 \% & 76.9 \% (*)&  7.6 \% (*) \\
		\hline
		\hline
		top. AOS & \centering 3f & 99.8 \% &  97.9 \% & 80.2 \% &  1.7 \%  \\
		\hline
		AzEl & \centering 3f & 99.8 \% &  98.0 \% & 79.2 \% &  1.8 \%  \\
		\hline
		XYZ & \centering 3f & 99.7 \% &  97.8 \% & 81.3 \% &  1.8 \%  \\
		\hline
	\end{tabular}
	\label{tab:res}
\end{table}

\subsubsection{Estimation Bias}
\label{sec:bias}

\begin{figure*}[htbp]
	\centering
	\input{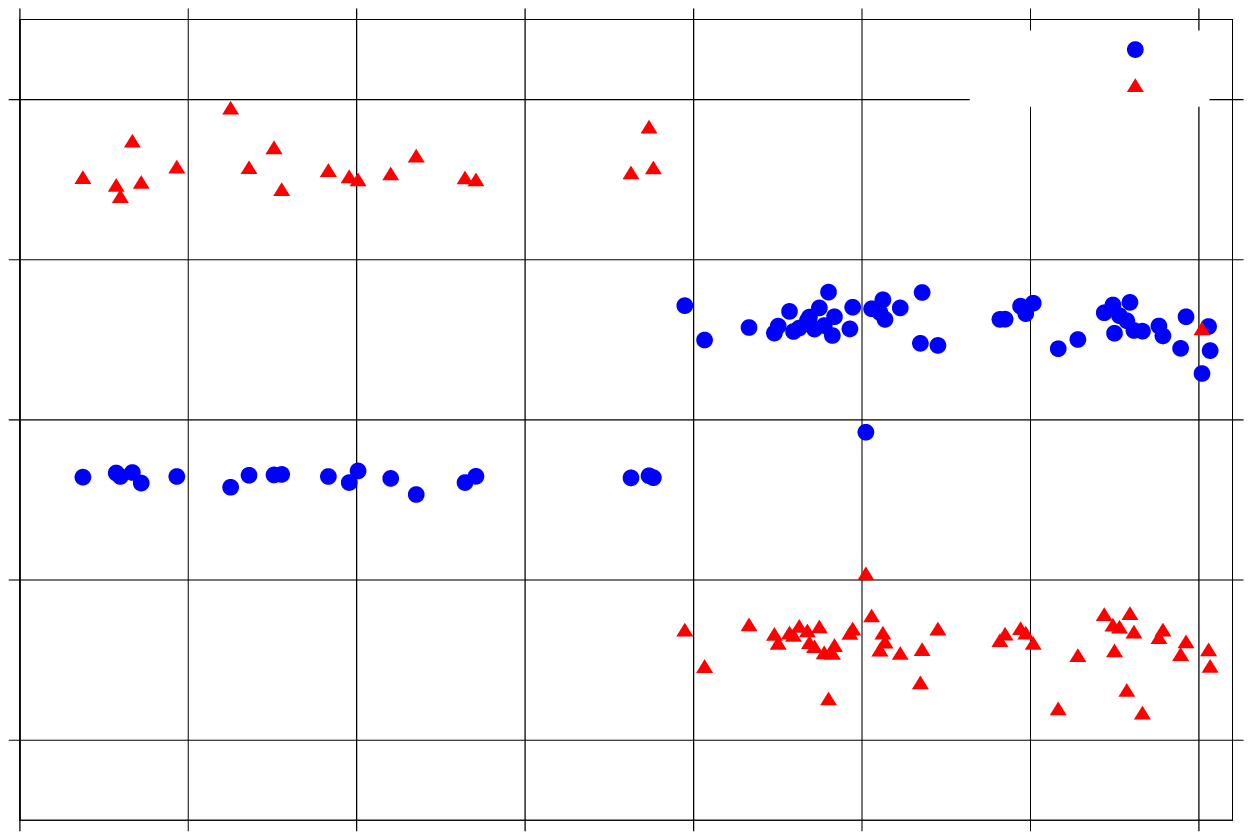}%
	\caption{Distribution of discriminators $\dot{\rho}_1$ and $\Delta a$ for the 3f-survey over the RAAN using objects on polar orbits with incomplete revolutions.}
	\label{fig:deltarhodots}
\end{figure*}

In the results shown, the distribution of the Mahalanobis distances appeared to be shifted compared to the expected distribution which is a $\chi$-distribution with two degrees of freedom due to the two-dimensional discriminator vector under the assumption that the values of the discriminators are normally distributed. In case of this south-staring survey campaign, the normality of the discriminators breaks down because of the measurement geometry and short-periodic perturbations. Especially objects on polar orbits are detected once on their ascending and once on their descending arc within 24 hours, thus they do not exhibit an approx. integer multiple of revolutions between detections. The applied J$_2$-correction of the mean motion does only correct the perturbed motion for a full revolution, thus incomplete revolutions are estimated with a bias in the semi-major axis. This effect can be seen in \Fig{fig:deltarhodots}, which relates the biases in the semi-major axis to the RAAN for polar orbits. For one day of measurements from a single station, the RAAN is related to the direction of motion (ascending or descending arc at first detection), which can be seen via the range-rate $\dot{\rho}_1$ at the first pass. For a south-staring FoR, a negative range-rate (approaching) is caused by an object on the ascending part of a polar orbit. Depending on the direction of motion, different short-periodic perturbations during the incomplete revolution lead to different biases in the semi-major axis.

Combining this bias, which influences especially the velocity, with the measurement geometry of observing either near-maximum positive or negative range-rates on the descending or ascending arc, respectively, leads to an offset of the discriminator values to opposite, positive and negative, directions. Thus, there is no normal distribution any more and the errors are larger than expected from the measurement statistics. If the discriminator $\Delta \dot{\rho}$ is split into one part for the measurement uncertainty $\Delta\dot{\rho}_M$ and one part due the bias of the semi-major axis $\Delta\dot{\rho}_a$, the Mahalanobis distances become shifted towards larger values because the estimated uncertainty $C_M$ only refers to the measurement:
\begin{eqnarray}
M_d^2 &=& \left( \Delta\dot{\rho}_M + \Delta\dot{\rho}_a   \right)^T \cdot C_M^{-1} \cdot \left( \Delta\dot{\rho}_M + \Delta\dot{\rho}_a   \right) \  \\
 &=& \Delta\dot{\rho}_M^T \cdot C_M^{-1} \cdot  \Delta\dot{\rho}_M + 2 \cdot \Delta\dot{\rho}_a^T \cdot C_M^{-1} \cdot  \Delta\dot{\rho}_M + \Delta\dot{\rho}_a^T \cdot C_M^{-1} \cdot  \Delta\dot{\rho}_a \ .
 \label{eq:mdbias}
\end{eqnarray}
In \Equ{eq:mdbias}, only the first term is $\chi^2$-distributed, while the remaining terms including $\Delta\dot{\rho}_a$ are introducing the observed bias, which is independent of the measurement uncertainty. The larger the uncertainty in $C_M$, the smaller the effect of the bias terms because $\Delta\dot{\rho}_a$ is independent of the measurement noise. Because of that, the distribution shifts closer to expected values for the experiments with larger detection intervals, which have higher absolute uncertainties.

\section{Conclusion}

In this paper, we presented and analysed several new approaches to use attributables for radar surveys. A new coordinate system with two different realisations was introduced to improve the fitting process of radar attributables. The comparison between four coordinate system has been made concerning the accuracy of the fitting process, which showed the improvement of the new coordinates compared to classical systems. From this first set of experiments, it was possible to derive rules for the fitting of the attributables concerning the order of the used polynomial depending on the observable and the tracklet length. It was shown that certain observables require up to sixth order polynomials for tracklets of up to 3 minutes length.

As a second step, the influence of different parameters, e.g. the length of the tracklet, the detection frequency, and the observations' noise level, during the attributable fitting on the correlation results was analysed. The results showed no cases with significant reductions in correlation performance, which also verifies the robustness of the applied methods. Also here, the new coordinate system, especially in the geocentric realisation, showed the best results with regard to correlation and orbit accuracy. It was also shown how longer tracklets improve the correlation result, thus the fitting process is able to transport the information gain via the attributable to the correlation level.

Finally, a surveillance radar scenario using a scanned Field of Regard was applied to test the entire process under more realistic conditions using different detection frequencies. Two additional filters, one checking the tracklet residuals and one with a least squares orbit determination, were introduced to reduce the number of false positives, which can be very high after the initial correlation step. Depending on the acceptance thresholds, the process usually identified more than 75 \% of the true correlations, while having only approximately 2 \% of false correlations and up to 8 \% false correlations for cases with a low number of detections and a higher acceptance threshold during the orbit determination. In this context, it was also shown how a bias in the J$_2$-perturbed initial orbit determination leads to a shift in the distribution of Mahalanobis distances. The successful correlation for all different measurement frequencies and coordinate systems shows how the attributable approach can be applied to the processing of surveillance radar data. Future work can include the treatment of the residuals' statistical correlation during the polynomial fit in the new coordinate systems and the application to real radar data.

\section*{Acknowledgements}
The first author is supported by the European Space Agency through the Networking/Partnering Initiative.

\printbibliography

\end{document}